\documentclass[lettersize,journal]{IEEEtran}
\usepackage[caption=false,font=footnotesize]{subfig}
\usepackage{graphicx}
\usepackage{url}
\usepackage{xcolor}
\usepackage{diagbox}
\usepackage{threeparttable}
\usepackage{booktabs}
\usepackage[nocompress]{cite}
\usepackage{amsfonts}
\usepackage{amsmath}
\usepackage{algorithm}
\usepackage{algpseudocode}

\begin{document}

	\title{CSI2Dig: Recovering Digit Content from Smartphone Loudspeakers \\ Using Channel State Information}		
	\author{Yangyang Gu, Xianglong Li, Haolin Wu, Jing Chen, Kun He, Ruiying Du, Cong Wu}		
	\maketitle
		
	\begin{abstract}
		Eavesdropping on sounds emitted by mobile device loudspeakers can capture sensitive digital information, such as SMS verification codes, credit card numbers, and withdrawal passwords,
		which poses significant security risks. 
		Existing schemes either require expensive specialized equipment, rely on spyware, or are limited to close-range signal acquisition.
		In this paper, we propose a scheme, CSI2Dig, for recovering digit content from Channel State Information (CSI) when digits are played through a smartphone loudspeaker.
		We observe that the electromagnetic interference caused by the audio signals from the loudspeaker affects the WiFi signals emitted by the phone's WiFi antenna.
		Building upon contrastive learning and denoising autoencoders,
		we develop a two-branch autoencoder network designed to amplify the impact of this electromagnetic interference on CSI.
		For feature extraction, we introduce the TS-Net, a model that captures relevant features from both the temporal and spatial dimensions of the CSI data.
		We evaluate our scheme across various devices, distances, volumes, and other settings.
		Experimental results demonstrate that our scheme can achieve an accuracy of 72.97\%.
	\end{abstract}
		
		
	\section{Introduction}
	Voice assistant services, such as voice broadcasting, have greatly facilitated access to information on mobile devices.
	However, the eavesdropping of key digital information, such as SMS verification codes, credit card numbers, and withdrawal passwords, can lead to significant financial losses for users.
	This potential for substantial harm has made acoustic eavesdropping on mobile devices a hot research topic.

	Existing acoustic eavesdropping techniques typically focus on extracting vibration information from motion sensors \cite{Spearphone, ba2020learning, gyrophone}, optical sensors \cite{2200fps, 277188, lidarphone, lidar}, and RF signals such as millimeter waves \cite{mmEve, mmphase, mmecho}, RFID \cite{9355596, RFspy, 10.1145/3494975},	WiFi \cite{WiVib}, and electromagnetic radiation \cite{chen_eavesdropping_2024, liao_eavesdropping_2024, TEMPEST, magear},
	which are then used to recover loudspeaker sound information.
	For example, Davis et al. \cite{2200fps} developed a method to capture micro-vibrations of objects near a loudspeaker using a high-speed camera, thereby recovering the sound.
	Wei et al. \cite{WiVib} leveraged software-defined radio to capture vibration-induced changes in WiFi signals and recover sound from loudspeakers.
	Periscope \cite{chen_eavesdropping_2024} uses electromagnetic leakage from an amplifier in the speaker to recover sound and employs a WiFi module on a development board for remote eavesdropping by transmitting sensor data to an analysis device.
	However, these methods either require expensive specialized equipment or rely on close-range signal acquisition, which increases both the cost of eavesdropping and the risk of exposure.
	Inspired by the work of Periscope, we investigate whether it is possible to directly recover digit sequences played by a smartphone loudspeaker from the WiFi signals emitted by the device.

	WiFi, being a form of electromagnetic wave, is generated by the synchronous oscillation of electric and magnetic fields.
	Half of the energy in electromagnetic waves is contained in the electric field,
	while the other half is in the magnetic field \cite{halliday2013fundamentals}.
	Thus, the magnetic field generated by the speaker while playing audio may influence the WiFi signals emitted by the device’s WiFi antenna.
	Specifically, this effect could manifest in the Channel State Information (CSI),
	which is sensitive to changes in the physical channel \cite{li_wifinger_2016, chen_rapid_2017}.

	To validate this hypothesis, we conduct a preliminary experiment.
	In this experiment, we use a smartphone (Honor V30 Pro) as the target device and collect CSI data while the device plays an audio file containing a single digit, as well as when no audio is played.
	The results reveal a significant difference in the CSI distribution between these two conditions.
	Moreover, the changes in CSI are temporally synchronized with the variations in the sound wave,
	suggesting that the loudspeaker does indeed affect the CSI.
	Additionally, we investigate the consistency of the effect of the same digit on CSI and the distinctiveness of different digits.
	The experimental results demonstrate that CSI fingerprints for each digit can be used to recover them effectively.
	Detailed experimental setup and results are provided in Section \ref{sec:3}.

	To extract meaningful information from CSI, we employ a two-branch autoencoder network according to the principles of supervised contrastive learning \cite{SCL} and denoising autoencoders \cite{vincent2010stacked}.
	The purposes of the network is to make input samples with the same label output more similar generated samples, while input samples with different labels output more different generated samples.
	For feature extraction, we design the TS-Net model to capture features from both temporal and spatial dimensions.
	In the temporal dimension, we use a Long Short-Term Memory (LSTM) network \cite{graves2012long},
	which is effective for sequential feature extraction, to capture the temporal variations of CSI subcarriers.
	The LSTM network processes each time step in the time series and optimizes the retention of information at each step during training.
	In the spatial dimension, we employ a Convolutional Neural Network (CNN) \cite{gu2018recent},
	which excels at extracting structural features from sequences, to process the CSI data.
	The spatial dimension refers to the distribution of subcarriers across different time points,
	which differs from the lateral variation in the temporal dimension.
	Our experiments, as well as previous studies \cite{rttwd, wifileaks}, indicate that the distribution of subcarriers also carries information about channel changes.
	Finally, we fuse the temporal and spatial features into a final feature representation of specific digits via a weighted fusion module.

	In this paper, we propose a scheme, CSI2Dig, to recover specific digit content from CSI when digits are played by a smartphone loudspeaker.
	Our scheme follows these steps.
	First, we remove subcarriers with no temporal variation or excessive noise based on observations and the official documentation of the CSI extraction tool.
	Next, we segment the CSI sequence into samples that contain digit information.
	We then reconstruct the CSI subcarriers using the two-branch autoencoder network to enhance the impact of electromagnetic interference on CSI.
	Finally, we classify the CSI samples into specific digits using the TS-Net model based on LSTM and CNN.
	This model extracts digit-related features from both the temporal and spatial dimensions and integrates them into a final feature representation.

The contributions of this paper are summarized as follows.
\begin{itemize}
  	\item We design a scheme, CSI2Dig, to recover digit content played by smartphone loudspeakers based on CSI.
	CSI2Dig identifies digits by establishing CSI fingerprints corresponding to the electromagnetic radiation effects on the WiFi antenna when the loudspeaker plays digits.
	\item We design a deep learning model TS-Net based on LSTM and CNN to extract features from both temporal and spatial dimensions and establish CSI fingerprints for digits.
	\item We evaluate the performance of CSI2Dig using three different smartphones at various distances and volumes.
	We can achieve an average accuracy of 58.4\% at four meters using only CSI data.
\end{itemize}

The rest of this paper is organized as follows.
Section \ref{sec:2} introduces the preliminaries and motivation of our work.
Section \ref{sec:3} presents the design of our scheme in detail.
Section \ref{sec:4} describes the implementation of our scheme and the comprehensive evaluation results.
Moreover, we discuss the limitations, experimental findings and future work in Section \ref{sec:5}.
Finally, related work is briefly captured in Section \ref{sec:6}, and a conclusion is drawn in Section \ref{sec:7}.


\section{Preliminaries}
\label{sec:2}
In this section, we first discuss the influence of Electromagnetic Radiations (EMR) on CSI, the relationship between the audio playback and CSI, and CSI amplitude distinctiveness for different digits.

\subsection{Influence of EMR on CSI}
This section discusses the influence of EMR on Channel State Information.
CSI is employed in the 802.11.x protocol, which uses orthogonal frequency division multiplexing (OFDM) to assess the physical state of the communication channel and improve signal propagation \cite{wifileaks}.
CSI can be estimated based on a known preamble transmitted between the transmitter and receiver \cite{preamble}.
A received CSI measurement, denoted as $\mathcal{H}$, is represented as:
\begin{equation}
\mathcal{H} =(\mathnormal{H(f_1)}, \cdots ,H(f_n)),
\end{equation}
where $n$ is the number of orthogonal frequencies (i.e., subcarriers) and $H(f_n)$ can be denoted as $H(f_n)=\vert H(f_n)\vert e^{i\angle H(f_n)}$ with the amplitude $\vert H(f_n)\vert$ and the phase $\angle H(f_n)$.

CSI is capable of capturing fine-grained variations at the physical layer, reflecting characteristics such as multipath propagation, including reflection, refraction, and diffraction of signals in the physical space \cite{survey}.
Additionally, CSI is influenced by factors like channel contention and imperfections in device hardware \cite{li_wifinger_2016, chen_rapid_2017, WiAnti}.
One aspect that has not been sufficiently emphasized in existing literature is the impact of electromagnetic interference (EMI) due to imperfect hardware design.

As modern electronic devices become increasingly integrated, the potential for electromagnetic interference between different components is unavoidable \cite{EMI}.
While manufacturers have implemented electromagnetic shielding measures to mitigate this interference,
these techniques can only reduce, but not entirely eliminate EMI \cite{chen_eavesdropping_2024}.
Of particular concern is the electromagnetic radiation generated by circuit switching,
which can have a significant impact on nearby electronic devices.

In this context, we focus on the electromagnetic radiation emitted by loudspeakers and its influence on WiFi signal transmission.
As discussed in \cite{chen_eavesdropping_2024}, when a loudspeaker emits sound, it inevitably generates electromagnetic radiation.
This radiation alters the magnetic field in the surrounding area, which in turn affects the electromagnetic waves emitted by the WiFi antenna.
WiFi signals, which consist of continuously oscillating in-phase electric and magnetic fields \cite{Griffiths2023}, are particularly susceptible to these changes.
The energy of an electromagnetic wave, which is derived equally from the electric and magnetic fields \cite{halliday2013fundamentals}, can be expressed as:
\begin{equation}
  \label{eq:u}
  u = \frac{1}{2\mu _0}B^2 + \frac{\epsilon _0}{2}E^2 , 
\end{equation}
where $E$ and $B$ denote the values of the electric and magnetic fields, respectively, and $\mu _0$ and $\epsilon _0$ denote the electric and magnetic constants, respectively.
CSI can be sensitive to changes in the electromagnetic radiation of the channel.

\begin{figure}[!t]
  \centering
  \includegraphics[width=0.95\linewidth]{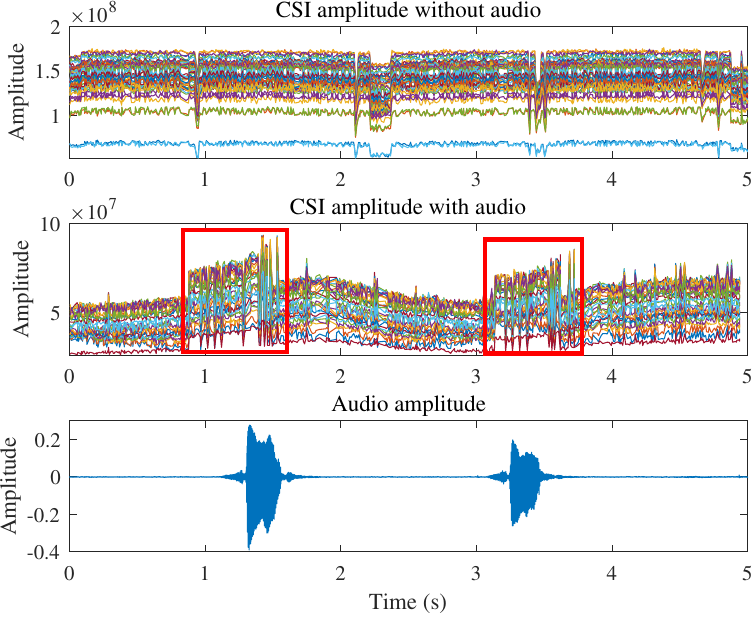}
  \caption{The top two sub-figures show the amplitude of CSI subcarriers without and with audio playback. The bottom sub-figure shows the audio amplitude of the digit SEVEN.}
  \label{fig:csiempsev}
\end{figure}

\subsection{Relationship between the Audio Playback and CSI}
\label{sec:2.2}
This section explores the correlation between audio playback and CSI amplitude variations through both theoretical analysis and experimental validation.
As demonstrated in \cite{chen_eavesdropping_2024}, electromagnetic leakage from a loudspeaker's amplifier correlates with the audio content, enabling the recovery of the audio based on this leakage.
According to Eq. \ref{eq:u}, the electromagnetic field changes caused by this leakage may affect the WiFi signal emission, which can be captured by CSI.
To verify this hypothesis, we conducted a simple experiment to investigate the relationship between audio playback and CSI changes.

We placed a smartphone flat on a table as the target device, playing a recorded audio file.
The audio consists of a volunteer’s voice, recorded with a smartphone’s built-in recorder app, pronouncing the digit 7 approximately once every two seconds.
A Nexus 6P phone, equipped with the Nexmon\_csi project \cite{nexmoncsi}, acted as the monitoring device, collecting CSI data packets from the target device based on its MAC address.
The devices were positioned one meter apart on separate tables to prevent solid vibrations from affecting the signal.
To minimize interference from other smartphone applications, the target device was running no background apps during audio playback.
To ensure sufficient WiFi packet transmission for CSI collection, a laptop on the same LAN sent ICMP packets to the target device at a frequency of 100Hz while it played the audio.
We collected CSI data both when the target device played audio and when it did not, for comparison.

The upper sub-figure of Figure \ref{fig:csiempsev} shows the amplitude changes of selected CSI measurements when the target device was not playing audio.
In this case, the CSI amplitudes remain generally stable, with no significant changes, consistent with observations from previous studies in stationary environments \cite{wifileaks}.
The middle sub-figure presents CSI measurements collected when the target device was playing audio.
Here, the red boxes highlight two instances of significant amplitude changes.
Due to limitations in firmware and other network factors, the monitor received fewer than 100 packets per second, resulting in packet loss at the 5-second mark, as indicated by the blank space in the sub-figure.
The lower sub-figure shows the amplitude of the played audio in the time domain.
The two marked amplitude changes correlate with the two digits played.

These amplitude changes in the CSI measurements and the audio playback are clearly time-correlated.
Although there is a slight timing deviation between the start of the audio playback and the CSI data collection (due to manual synchronization),
the experimental results clearly indicate that audio playback on the smartphone affects CSI amplitudes, providing a foundation for recovering digit content from CSI amplitude variations.

Furthermore, to investigate whether it is possible to distinguish whether the target device is playing audio,
we examine the differences between the two states (audio playing vs. audio not playing).
As observed in the upper two sub-figure of Figure \ref{fig:csiempsev},
there is a noticeable difference in the temporal change of the CSI amplitude between these two states.
To quantify these differences, we use correlation analysis in both the temporal and spatial dimensions, inspired by the work in \cite{loccams}.

In the temporal dimension, we measure the correlation between different subcarriers within the CSI sequence.
In the spatial dimension, we assess the correlation between the subcarrier distributions across different CSI measurements within the sequence. 
Specifically, for a CSI sequence $S = [\mathcal{H}_1, \mathcal{H}_2, \cdots , \mathcal{H}_w]^T$,
, where $w$ represents the number of CSI measurements.
We denote the sequence as $S = [H_1,H_2,\cdots, H_z]$, where $H_z$ represent the $z_{th}$ subcarrier.
The calculation of the two correlation matrices $\mathcal{C}_t$ and $\mathcal{C}_s$ is as follows: 
\begin{equation}
  \label{eq:corr1}
	\mathcal{C}_t(i,j)  = corr(H_i,H_j),
\end{equation}
\begin{equation}
  \label{eq:corr2}
  \mathcal{C}_s(i,j) = corr(\mathcal{H}_i,\mathcal{H}_j),
\end{equation}
where $\mathcal{C}_t(i,j)$ refers to the correlation coefficient of the $i_{th}$ subcarrier and $j_{th}$ subcarrier,
and $\mathcal{C}_s(i,j)$ refers to the correlation coefficient of the $i_{th}$ CSI measurement and $j_{th}$ CSI measurement.

\begin{figure}[!t]
	\centering
	\subfloat[Temporal dimension]{\includegraphics[width=0.47\linewidth]{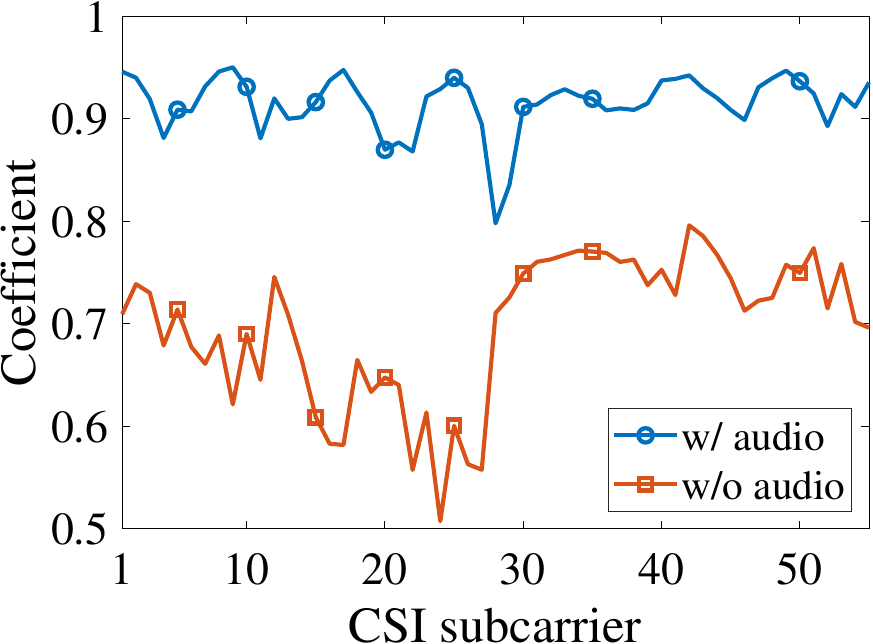}}
	\hspace{0.1em}
	\subfloat[Spatial dimension]{\includegraphics[width=0.47\linewidth]{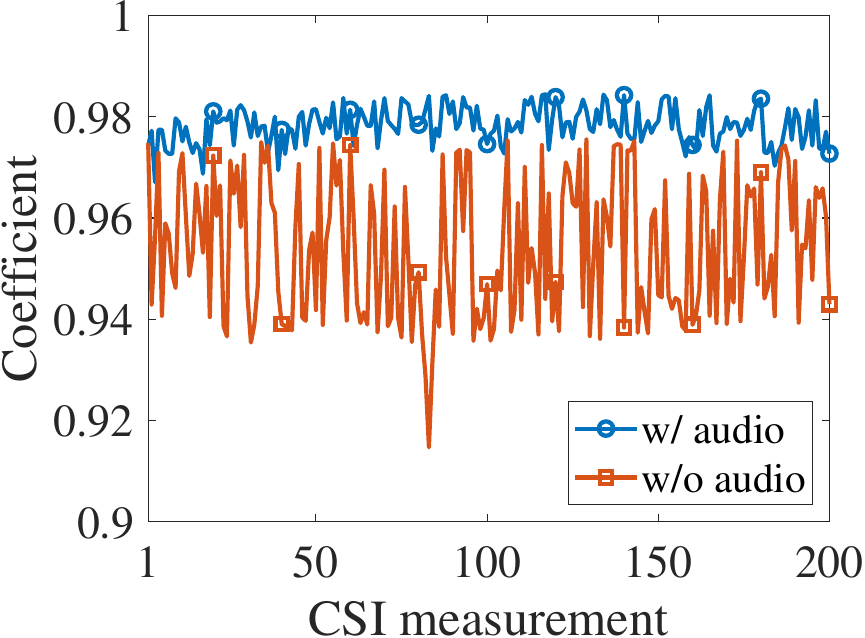}}
	\caption{Correlation coefficients of CSI sequences with/without audio playing in (a) temporal dimension and (b) spatial dimension.}
	\label{fig:corrtimespace}
\end{figure}

We first compute the correlation matrices for both the temporal and spatial dimensions of the CSI series shown in Figure \ref{fig:csiempsev}.
Next, we average the matrices column-wise to obtain the average correlation coefficients, as shown in Figure \ref{fig:corrtimespace}.
The results show relatively high correlations in both the temporal and spatial dimensions when the target device plays audio.
The temporal correlation exhibits greater variation, indicating that different subcarriers are affected to different degrees by audio playback.
In contrast, the spatial correlation is more stable, suggesting that subcarriers maintain stronger temporal consistency.
The average correlation coefficients for the temporal dimension are 0.915 and 0.797,
while for the spatial dimension, they are 0.979 and 0.952.
The larger differentiation in temporal correlations further supports the potential for distinguishing whether the target device is playing audio based on the temporal correlation of the CSI data.

\subsection{CSI Amplitude Distinctiveness for Different Digits}
\label{sec:2c}
To evaluate the feasibility of using CSI amplitude variations to recover the played digit content,
we conducted an experiment with two audio segments containing the digits 7 and 8.
The setup used for the experiment follows the procedures outlined in the previous section.
We analyzed both the temporal similarity of CSI changes for the same digit and the differentiation of CSI changes between different digits.

\begin{figure}[!t]
  \centering
  \includegraphics[width=0.95\linewidth]{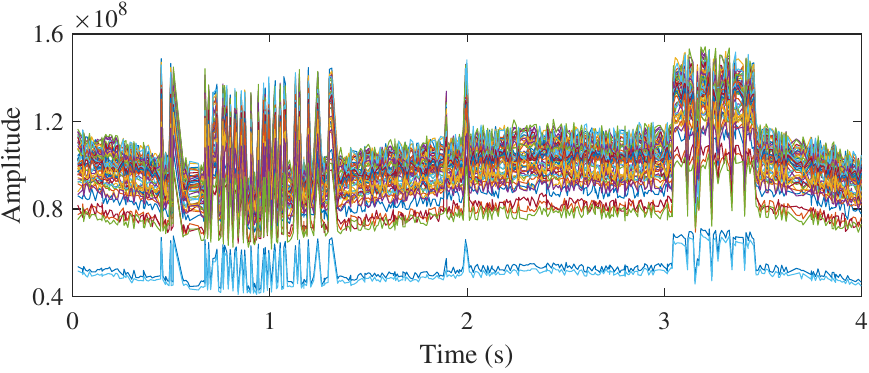}
  \caption{Amplitude of CSI subcarriers during playing the digit 8 twice.}
  \label{fig:csieig}
\end{figure}

We begin by analyzing the temporal dimension of the CSI changes.
The CSI amplitude variations caused by the playback of the digit 8 are shown in Figure \ref{fig:csieig},
while the middle sub-figure of Figure \ref{fig:csiempsev} illustrates the corresponding CSI changes for the digit 7.
From these, we can directly extract four CSI amplitude sequences: two sequences corresponding to the digit 7, denoted as $S^7 _1$ and $S^7 _2$ ,
and two sequences corresponding to the digit 8, denoted as $S^8 _1$ and $S^8 _2$.

To quantify the temporal similarity of the CSI changes for the same digit, we applied the Dynamic Time Warping (DTW) algorithm \cite{DTW} to measure the similarity between $S^7 _1$ and $S^7 _2$, and $S^8 _1$ and $S^8 _2$.
DTW is a widely used method for comparing audio waveforms.
The calculation is expressed as follows:
\begin{equation}
  d^i _{1,2} = dtw(S^i _1, S^i _2) \qquad i\in [7,8],
\end{equation}
where $d^i _{1,2}$ denotes the Euclidean distance between $S^i _1$ and $S^i _2$,
and $dtw(·)$ denotes the DTW algorithm.
To isolate the effect of temporal variations and remove the influence of CSI amplitude scale,
all sequences are normalized for each subcarrier prior to the calculation.
After normalization, each subcarrier has a mean of 1 and a standard deviation of 0.

To explore the ability to distinguish between different digits based on CSI amplitude variations,
we also applied the DTW algorithm to compute the similarity between CSI sequences from the digits 7 and 8, respectively. The calculation is as follows:
\begin{equation}
  d^{7,8} _{i,j} = dtw(S^7 _i, S^8 _j), \qquad i,j\in [1,2],
\end{equation}
where $d^{7,8} _{i,j}$ denotes the Euclidean distance between $S^7 _i$ and $S^8 _j$.
Since each sequence contains multiple subcarriers, the similarity of each subcarrier is calculated separately, and the average of all subcarriers is taken as the final similarity.

\begin{table}[!t]
   \centering
   \footnotesize
   \caption{The results of DTW and PCC for Different CSI Sequences from Different digits}
   \renewcommand\arraystretch{1.4}
   \begin{tabular}{m{1.4cm}<{\centering}m{1.1cm}<{\centering}m{1.1cm}<{\centering}m{1.1cm}<{\centering}m{1.1cm}<{\centering}}
      \toprule
      \diagbox{PCC}{DTW} & $S^7 _1$ & $S^7 _2$ & $S^8 _1$ & $S^8 _2$ \\
      \midrule
      $S^7 _1$ & - &27.22 & 54.60 & 40.51 \\
      $S^7 _2$ & 0.886 & - & 53.09 & 35.00 \\
      $S^8 _1$ & 0.774 & 0.754 & - & 46.36 \\
      $S^8 _2$ & 0.679 & 0.662 & 0.979 & - \\
      \bottomrule
   \end{tabular}
   \label{tab:dtwpcc}
\end{table}

The results of the DTW calculations are summarized in Table \ref{tab:dtwpcc}.
Our initial expectation is that the changes in CSI caused by the same digit would exhibit greater similarity in the temporal dimension, while the changes caused by different digits would differ more significantly. 
Therefore, we anticipate that the values of $d^i _{1,2}$ are smaller than $d^{7,8} _{i,j}$.
However, the results do not align with this expectation.
Specifically, the distances $d^7 _{1,2}$ and $d^8 _{1,2}$ are 27.22 and 46.36,
and the distances $d^{7,8} _{1,1}$, $d^{7,8} _{1,2}$, $d^{7,8} _{2,1}$, and $d^{7,8} _{2,2}$ are 54.6, 40.51, 53.09, and 35.00, respectively.
These results suggest that it is not feasible to distinguish different digits solely based on the temporal variations in CSI amplitude.

Inspired by previous works \cite{wifileaks,rttwd},
we next examine the spatial correlation of CSI measurements to explore the observed differences. 
Specifically, we use the Pearson Correlation Coefficient (PCC) \cite{PCC} to measure the similarity between CSI measurements across the four sequences $S^7 _1$, $S^7 _2$, $S^8 _1$, and $S^8 _2$.
As shown in Eq.\ref{eq:corr2}, we first calculate the correlations of CSI measurements affected by the same digits.
The mean correlation coefficients for $S^7 _1$ and $S^7 _2$, and for $S^8 _1$ and $S^8 _2$, denoted as $c^7 _{1,2}$ and $c^8 _{1,2}$ respectively, are calculated as follows:
\begin{equation}
    c^i _{1,2} = \frac{1}{N_1 N_2}\sum_{p = 1}^{N_1}\sum_{q = 1}^{N_2}corr(\mathcal{H}^i _{1,p},\mathcal{H}^i _{2,q}), i\in [7,8],
    \label{eq:cii}   
\end{equation}
where $N_1$ and $N_2$ denotes the CSI measurement number of the sequences $S^i _1$ and $S^i _2$,
and $corr(\mathcal{H}^i _{1,p},\mathcal{H}^i _{2,q})$ represents the correlation coefficient between the $p_{th}$ CSI measurement in $S^i _1$ and the $q_{th}$ CSI measurement in $S^i _2$.
Next, we calculate the average correlation coefficients between CSI measurements in sequences of different digits.
\begin{equation}
  c^{7,8} _{i,j} = \frac{1}{N_3 N_4}\sum_{p = 1}^{N_3}\sum_{q = 1}^{N_4}corr(\mathcal{H}^7 _{i,p},\mathcal{H}^8 _{j,q}), i,j\in [1,2],
  \label{eq:cij}   
\end{equation}
where $N_3$ and $N_4$ denotes the CSI measurement number of the sequences $S^7 _i$ and $S^8 _j$,
and $corr(\mathcal{H}^7 _{i,p},\mathcal{H}^8 _{j,q})$ represents the correlation coefficient between the $p_{th}$ CSI measurement in $S^7 _i$ and the $q_{th}$ CSI measurement in $S^8 _j$.
The final calculation results are shown in Table \ref{tab:dtwpcc}.
In contrast to the temporal correlation, the spatial correlation results align with our expectations.
The mean correlation coefficients for $S^7 _1$ and $S^7 _2$, and for $S^8 _1$ and $S^8 _2$ are 0.886 and 0.979,
while the values of $c^{7,8} _{1,1}$, $c^{7,8} _{1,2}$, $c^{7,8} _{2,1}$, and $c^{7,8} _{2,2}$ are 0.774, 0.679, 0.754 and 0.662, respectively.
The effect of the same digit on the CSI subcarrier distribution is indeed more similar than the effect of different digits on the CSI distribution.   
This experimental results show that it is feasible to recover the played digit content from the perspective of CSI subcarrier distribution.

\begin{figure}[!t]
  \centering
  \includegraphics[width=0.95\linewidth]{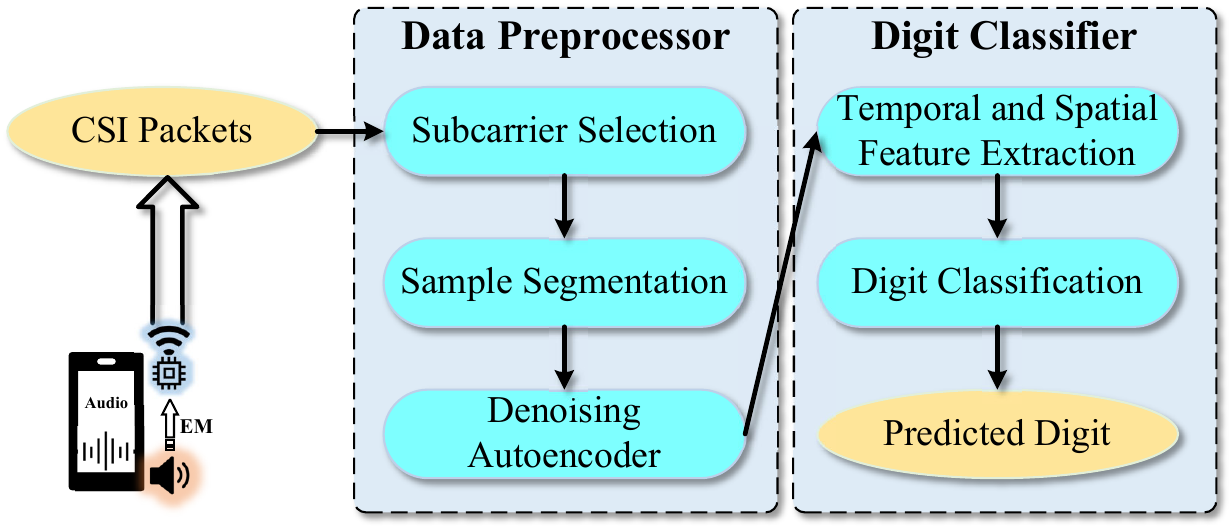}
  \caption{Workflow of CSI2Dig.}
  \label{fig:scheme}
\end{figure}

\section{System Design}
\label{sec:3}
\subsection{System Overview}
In this section, we describe the workflow of CSI2Dig.
As shown in Figure~\ref{fig:scheme}, CSI2Dig consists of two modules, \textit{Data preprocessor} and \textit{Digit Classifier}.
Data preprocessor module first implements subcarrier selection to remove those subcarriers that are functional and those that are clearly excessively noisy.
Next, a sample segmentation method is used to obtain samples of CSI sequences.
Finally, we employ a denoising autoencoder to reconstruct CSI subcarriers to enhance the impact caused by the loudspeaker.
Digit Classifier module contains a designed deep neural network \textit{TS-Net} to extract features from CSI amplitude samples from temporal and spatial dimensions, and classify samples to exact digits.

\subsection{Data Preprocessor}
\textit{Subcarrier selection.}
According to the protocol standard of OFDM and the official documentation of the CSI extraction tool,
we first remove the subcarriers that are used as guide and isolation bands in the subcarriers \cite{nexmoncsi,preamble}.
The variations of these subcarriers have a poor correlation with the surrounding subcarriers and fluctuate almost randomly.
Finally, we remove some subcarriers that always have a lot of burr noise to avoid interfering with the extraction of valid information in the subsequent modules based on our observation.
This step is to remove the subcarriers that obviously do not contain valid information for the digit recovery.

\begin{figure}[!t]
	\centering
	\subfloat[Raw CSI measurements]{\includegraphics[width=0.7\linewidth]{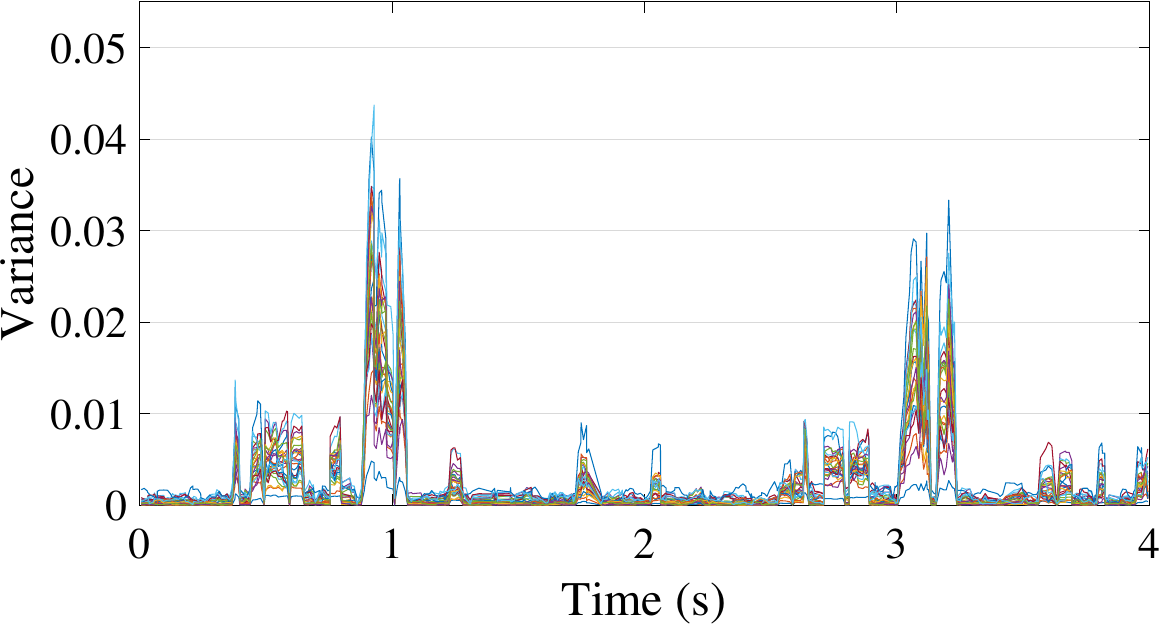}\label{fig:vara}}
	\vspace{0.3em}
	\subfloat[Denoised CSI measurements]{\includegraphics[width=0.7\linewidth]{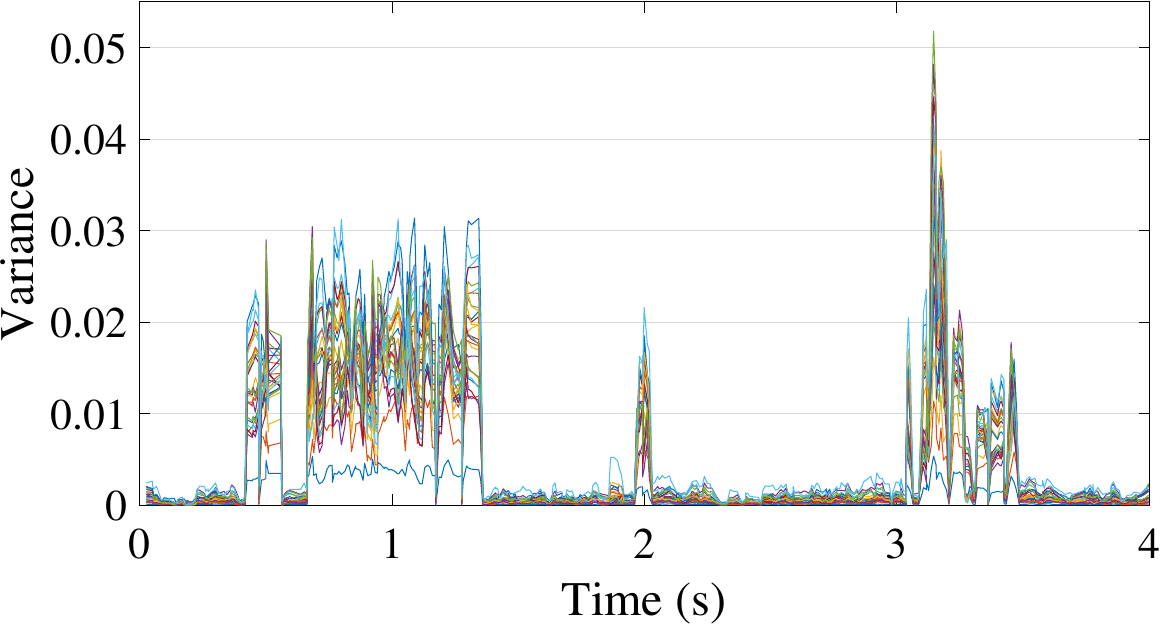}\label{fig:varb}}
	\caption{Variance of CSI measurements in (a) the Figure \ref{fig:csiempsev} and (b) the Figure \ref{fig:csieig}.}
	\label{fig:var}
\end{figure}

\begin{algorithm}[!t]
	
	\caption{Sample Segmentation}
	\renewcommand{\algorithmicrequire}{\textbf{Input:}}
	\renewcommand{\algorithmicensure}{\textbf{Output:}}
	\begin{algorithmic}[1]
	\Require CSI sequence with timestamps
	\Ensure Processed samples $R_{N_t \times N_s}$
	\State Segment the CSI sequence into samples every 2 seconds based on the timestamp
	\For{each sample}
		\State $N \gets$ number of CSI measurements in the sample
		\If{$N < \frac{N_{norm}}{2}$} 
			\State Discard the sample
		\Comment{$N_{norm}$ \textit{is the number of CSI measurements that should be collected in 2 seconds based on the packet sending rate}.}
		\ElsIf{$\frac{N_{norm}}{2} \leq N < N_{norm}$}
			\State Apply linear interpolation to restore $N$ to $N_{norm}$
		\Else{$N > N_{norm}$}
			\State Resample the sample to reduce $N$ to $N_{norm}$
		\EndIf
	\EndFor
	\State \Return $R_{N_t \times N_s}$
	\Comment{$N_t$ \textit{and} $N_s$ \textit{are the numbers of CSI measurements and CSI subcarriers}}
	\end{algorithmic}
	\label{alg1}
	\end{algorithm}
\textit{Sample segmentation.}
As shown in Figure \ref{fig:csiempsev}, the electromagnetic interference generated by audio playback produces a spiking noise on the CSI amplitude,
whereas the change in the CSI during the silent period without sound is relatively smooth.
Therefore, we first segment the original CSI amplitude sequence into samples associated with each voice.
Based on the direct observation and empirical analysis of the data,
it is natural to think of utilizing the variance of the CSI subcarriers for sample segmentation.
We calculate the variance of the CSI sequences in the Figure \ref{fig:csiempsev} and Figure \ref{fig:csieig}.
However, the variance is not a reliable metric.
As shown in Figure \ref{fig:var}(a), there is still a noticeable variation in variance when the loudspeaker is not playing audio due to the presence of interference.
This situation can misjudge the start point of the sample sequence and lead to errors in the sample segmentation.
Moreover, the change in CSI variance caused by different audio is variable.
The variance onset points in Figure \ref{fig:var}(a) and Figure \ref{fig:var}(b) are around 0.01 and 0.02, respectively.
Therefore, we cannot use the same variance threshold to segment different samples.

As discussed in Section \ref{sec:2.2}, the recorded audio sounds every two seconds.
Thus, we employ an algorithm to use an equal time interval method to segment the samples as shown in Algorithm \ref{alg1}.
One sample contains 2 seconds of CSI measurements with the number $N_{norm}$ according to the packet sending rate.
However, the frequency of packets received by the monitoring device is not stable due to the packet loss rate and the device's own WiFi packets.
Therefore, after segmenting a sample every two seconds based on the timestamp of the CSI sequence (Line 1), 
we examine the number of CSI measurements contained in the sample.
If this number is less than half of the normal number, we discard the sample (Line 4-5).
If the number is not less than half but less than the normal number, we sample a linear interpolation method to restore the number to the normal number (Line 6-7).
If the number is greater than normal number, we resample the sample to ensure that the number of CSI measurements is right (Line 8-9).
After this step, we can obtain a sample $R_{N_t \times N_s}$, where $N_t$ and $N_s$ are the numbers of CSI measurements and CSI subcarriers.

\begin{figure}[!t]
	\centering
	\subfloat[Raw CSI measurements]{\includegraphics[width=0.7\linewidth]{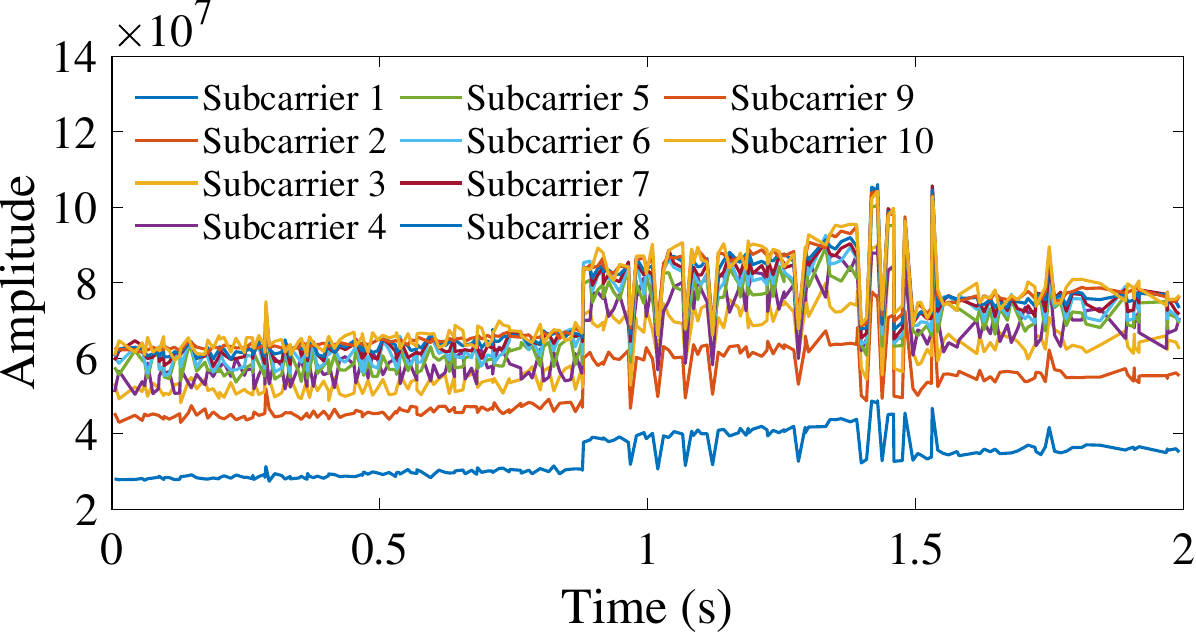}\label{fig:waveleta}}
	\vspace{0.5em}
	\subfloat[Denoised CSI measurements]{\includegraphics[width=0.7\linewidth]{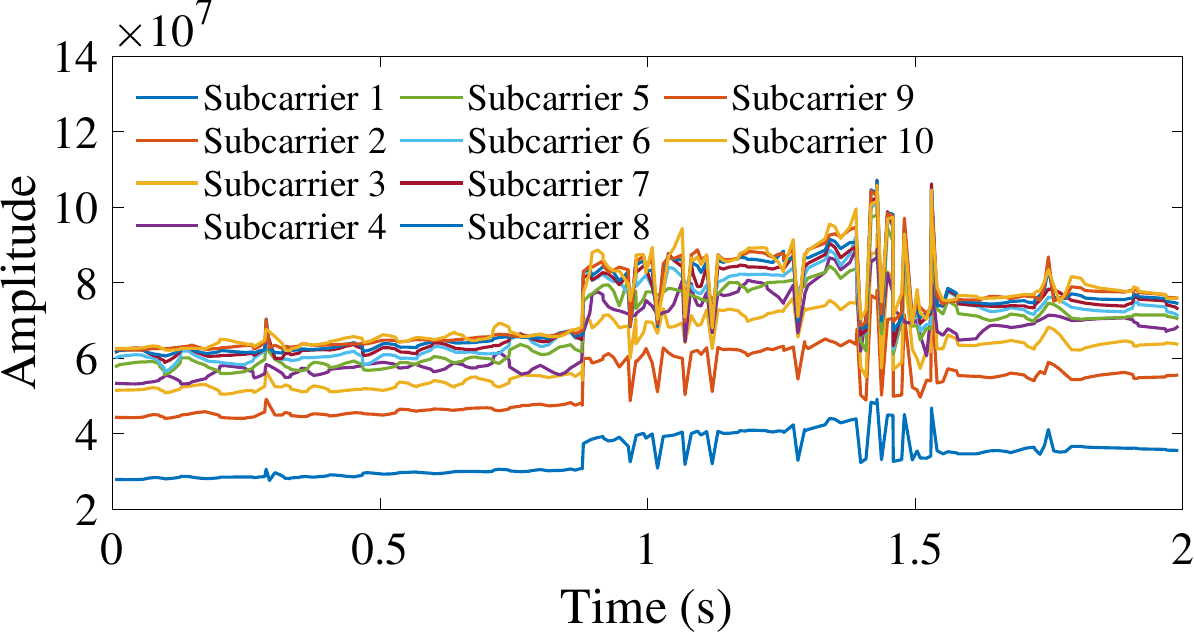}\label{fig:waveletb}}
	\caption{CSI measurements (a) before and (b) after applying wavelet denoising.}
	\label{fig:wavelet}
\end{figure}
\textit{Denoising autoencoder.}
The interference to CSI is varied and consists mainly of interference from the human body (even the stationary one \cite{wifileaks}) and channel attenuation.
We use the commonly used wavelet denoising method \cite{DWT} to preprocess the data.
As shown in Figure \ref{fig:wavelet}, the CSI amplitude becomes relatively smooth after using wavelet denoising.
Probably due to the loss of some details, the accuracy of the denoised samples is found to be slightly lower than the original samples during testing.
We try to modify the parameters of wavelet denoising and there is no significant improvement.
Therefore, we focus on a deep learning based denoising method, the denoising autoencoder.
Unlike the traditional autoencoder, we do not add noise to the raw data because the raw data itself contains noise.
Therefore, we develop a two-branch autoencoder neural network based on linear layers using the idea of contrastive learning, where the encoder and the decoder both contain three linear layers. 
Taking sample pairs as inputs, we use contrastive loss to reinforce the similarity of same-labeled data and the difference of different-labeled data.
Assuming that given a pair of samples ($x_i$,$x_j$) and their corresponding labels $y_{ij}$ (1 for the same sample pairs and 0 for the different sample pairs),
the object function $\mathcal{L}_{\text{c}}$ can be denoted as:
\begin{equation}
	\begin{split}
		\mathcal{L}_{\text{c}} = \Big[ & \frac{1}{2N_c} \sum_{i,j}  y_{ij} \, corr(x_i, x_j)^2 \\
& + (1 - y_{ij}) \, \max(\xi - corr(x_i, x_j), 0)^2 \Big],
	\end{split}
\end{equation}
where $N_c$, and $corr(x_i,x_j)$ represents the number of samples and the correlation coefficient between the sample $x_i$ and the sample $x_j$.
$\xi$ is the threshold for the correlation coefficient and we set it as 0.85 in our work.

\subsection{Digit Classifier}
This section will introduce the deep neural model which extracts features from CSI samples and classify the sample to the exact digit.

\begin{figure}[!t]
  \centering
  \includegraphics[width=0.95\linewidth]{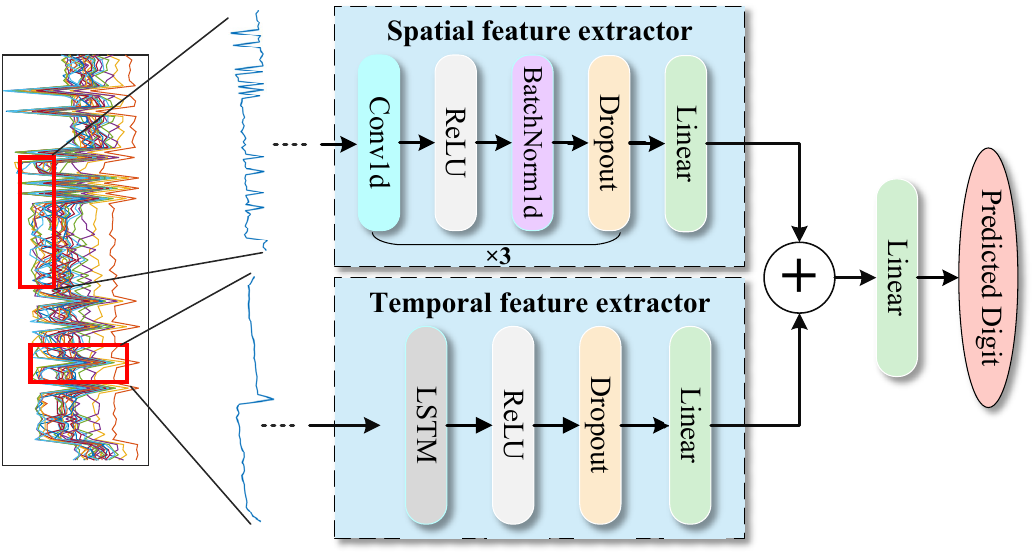}
  \caption{Structure of the TS-Net for extracting spatial and temporal features from CSI samples.}
  \label{fig:model}
\end{figure}
As shown in Figure \ref{fig:model}, we design a two-branch network to extract features from the temporal and spatial dimensions, respectively.
In the temporal dimension, each sample contains $N_s$ sequences of length $N_t$, i.e., CSI subcarriers.
The temporal variation of CSI amplitude is generated by playing continuous audio from a loudspeaker.
In other words, continuous variation in amplitude indicates a meaningful audio.
Therefore, we use a LSTM neural network to extract the temporal features of each subcarrier.
LSTM can efficiently extract timing features because of its unique cellular structure \cite{graves2012long}.
The LSTM cell contains a forgetting gate, an input gate, and an output gate.
The role of the forgetting gate is to decide how much information to discard away through the Sigmoid function $\sigma$, i.e., to selectively forget the information of the cell state in the previous step.
The working principle of the forgetting gate can be expressed as follows:
\begin{equation}
  \label{eq:ft}
  f_t = \sigma (W_f \cdot  [h_{t-1},\mathcal{H}_t] + b_f),
\end{equation}
where $\mathcal{H}_t$ and $h_{t-1}$ denote the CSI measurements of the current time step and the hidden state of the previous time step, and $W_f$ and $b_f$ denote the weight matrix and bias vector of the output $f_t$.
The forgetting gate generates a probability array $f_t$ in the interval [0,1] for the output $c_{t-1}$ of the previous time step.

The input gate determines what is stored in the current cell state, i.e., new information is selectively recorded into the cell state.
It contains two main activation layers, \textit{Sigmoid} and \textit{tanh}.
The Sigmoid layer $\sigma$ mainly outputs an array of probabilities $i_t$ that determines the retained information.
The tanh layer $tanh$ serves to output a candidate vector $\tilde{C}_t$, which will be added to the cell state.
The specific calculation is as follows:
\begin{equation}
  \label{eq:it}
  i_t = \sigma(W_i \cdot  [h_{t-1},\mathcal{H}_t] + b_i),
\end{equation}
\begin{equation}
  \label{eq:ctt}
  \tilde{C}_t = tanh(W_C \cdot  [h_{t-1},\mathcal{H}_t] + b_C),
\end{equation}
where $W_i$ and $b_i$, $W_C$ and $b_C$ denote the weight matrix and bias vector of $i_t$ and $tilde{C}_t$, respectively.
After inputting the gate, the cell state $C_t$ of the current time step can be updated directly.
The calculation is as follows:
\begin{equation}
  \label{eq:ct}
  C_t = f_t \ast C_{t-1} + i_t \ast \tilde{C}_t ,
\end{equation}
where $\ast $ denotes the operation of multiplying the corresponding elements instead of matrix multiplication.
As can be seen from Eq.\ref{eq:ct}, the cell state $C_t$ contains both the output of the previous time step $C_{t-1}$ and the input information of this time step $\tilde{C}_t$,
which allows the final output of the LSTM to learn the information of the whole temporal sequence.

The output gate determines the final output characteristics of the sequence,
which also contains two activation layers, \textit{Sigmoid} and \textit{tanh}.
The Sigmoid layer outputs a probability array $o_t$ to selectively output information about the state $C_t$ of this cell.
This is computed as follows:
\begin{equation}
  \label{eq:ot}
  o_t = \sigma(W_o \cdot [h_{t-1},\mathcal{H}_t] + b_o),
\end{equation}
where $W_o$ and $b_o$ denote the corresponding weight matrix and and bias vector, respectively.
The tanh layer normalizes the value domain of the cell state ct to [-1,1] to avoid amplitude differences from adversely affecting the loss calculation.
The output of the current time step, i.e., the hidden state $h_t$, is computed as follows:
\begin{equation}
  \label{eq:ht}
  h_t = o_t \ast tanh(C_t).
\end{equation}
In model training, the shape of $h_t$ is (num\_layers, batch\_size, hidden\_size).
In our model, the number of layers of LSTM is only 1, i.e. num\_layers = 1.
Therefore, after removing the dimensions whose dimension size is only 1, the shape of the temporal feature $F_t$ is (batch\_size, hidden\_size).

In the spatial dimension, we design a feature extraction network based on the convolutional neural network.
As described in \cite{loccams}, the distribution of subcarriers also reflects the channel variations and is demonstrated by our experiments in Section \ref{sec:2c}.
Since our data is a one-dimensional sequence, we use a one-dimensional convolutional layer to extract features.
The one-dimensional convolutional neural network extracts local features by sliding a fixed-size convolutional kernel over the input data,
and progressively extracts more advanced features through multiple convolutional and pooling layers.
Specifically, for the input CSI subcarrier $H$  and convolution kernel $w$, the one-dimensional convolution is computed as follows:
\begin{equation}
  y(t) = (H\ast w)(t) = \sum_{k = 0}^{K-1}H(t+k)\cdot w(k),
\end{equation}
where $t$ is the time step and $K$ is the size of the convolution kernel.
Our model contains three one-dimensional convolutional layers.
Each convolutional layer is followed by a \textit{ReLU} activation layer, a \textit{BatchNorm} layer, and  a \textit{Dropout} layer to provide nonlinear features and prevent overfitting.
Finally, with a linear layer, the model maps the length of each output channel to one dimension.
After removing dimensions with a dimension size of one,
the shape of the spatial feature $F_s$ is (batch\_size, out\_channels).

\subsection{Loss Function}
After extracting features in the temporal and spatial dimensions, we fuse these two features.
Specifically, we assign different weights to the features of the two dimensions.
The final feature $F$ is computed as follows:
\begin{equation}
  \label{eq:f}
  F = \alpha F_t + \beta F_s,
\end{equation}
where $\alpha$ and $\beta$ are the coefficients of the temporal dimension feature and the spatial dimension feature, respectively.
Finally, the model maps the feature $F$ through a linear layer to a one-dimensional predictive label with dimensions equal to the number of categories.
For the true label of the data, we use one-hot coding.
In the training phase, we use the cross entropy loss function to optimize the model.
Specifically, our goal is to minimize the objective function $\mathcal{L}$:
\begin{equation}
  \label{eq:L}
  \mathcal{L} = -\frac{1}{M_s}\sum_{i = 1}^{M_s}\sum_{j=1}^{M_c} y_{i,j}log(\hat{y}_{i,j}),
\end{equation}
where $M_s$ and $M_c$ are the number of samples and classes respectively,
and  $ y_{i,j}$ denotes the one-hot coded value of the $i_{th}$ sample in the $j_{th}$ class,
and $\hat{y}_{i,j}$ denotes the probability that the $i_{th}$ sample is predicted to be in the $j_{th}$ class.

\section{Evaluation}
\label{sec:4}
In this section, we will introduce the detailed setup of our experiments and the performance of CSI2Dig under extensive scenarios and settings.

\begin{figure}[!t]
	\centering
	\includegraphics[width=0.65\linewidth]{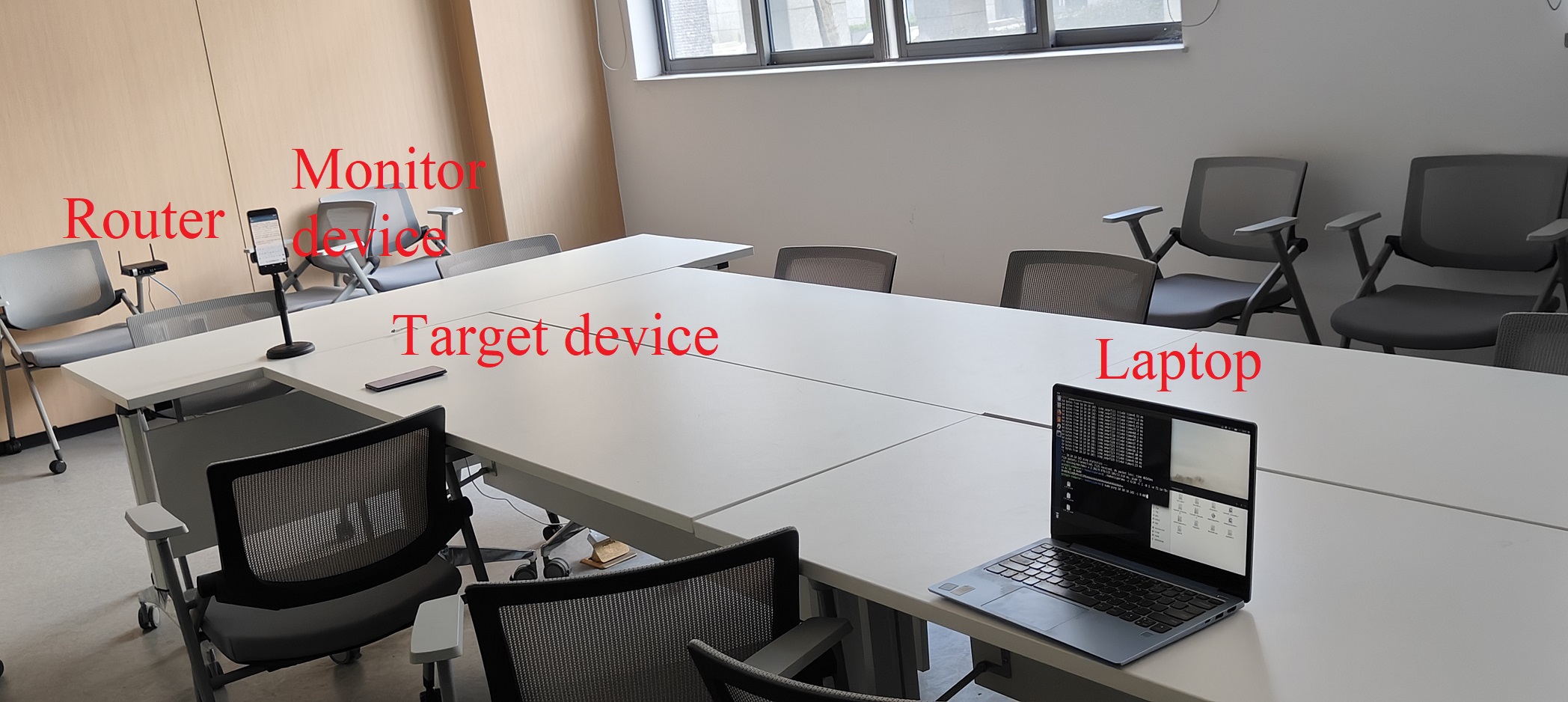}
    \caption{Discussing room used in our experiments.}
	\label{fig:room4}
\end{figure}
\begin{figure}[!t]
	\centering
	\includegraphics[width=0.5\linewidth]{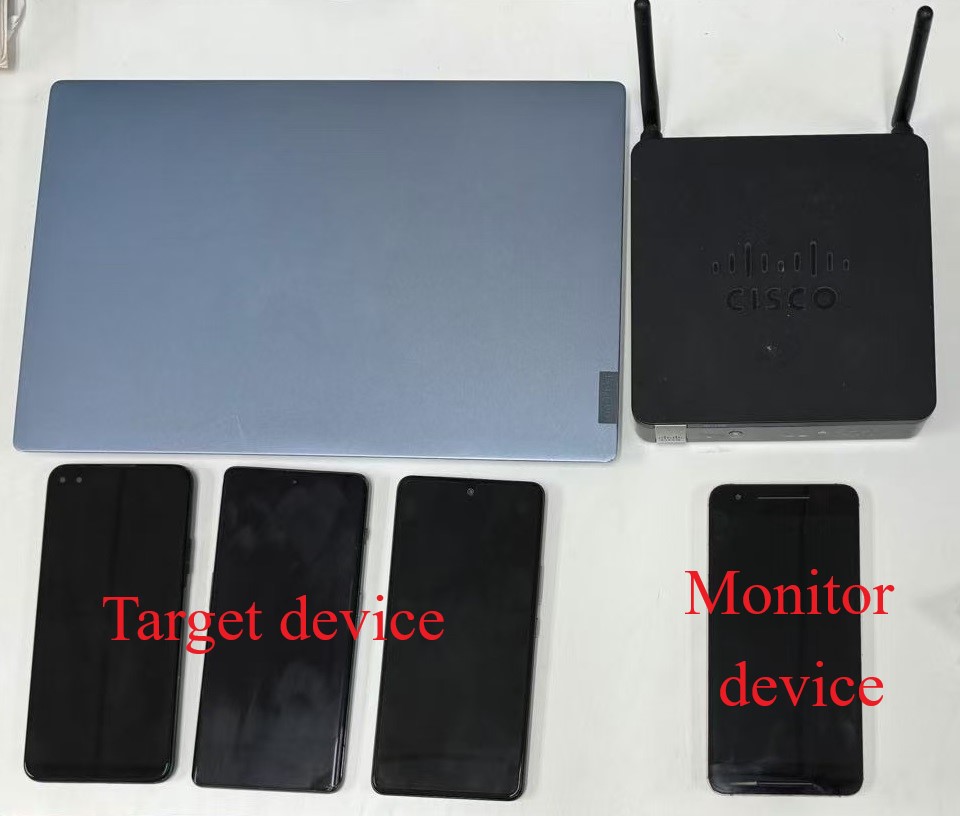}
    \caption{Devices used in our experiments.}
	\label{fig:dev}
\end{figure}
\subsection{Experimental Setup}
In this section, we will introduce the hardware and software in our experiments.
And the environment of collecting dataset and the default setup are also in this section.
Our data collection is conducted in a discussing room as shown in Figure \ref{fig:room4}.
We select three smartphones from different brands as target devices in the experiment.
The target devices in Figure \ref{fig:dev} are, from left to right, Honor V30 pro (Device 1), One Plus Ace 2 pro (Device 2) and IQOO neo 6se (Device 3).
With the help of the nexmon\_csi project \cite{nexmoncsi},
we use a smartphone Nexus 6P as the monitor device to collect CSI data.
Its network interference card is set to monitor mode to collect packets from WiFi devices based on their MAC addresses.
A cisco router is used as a WiFi access point to provide a WiFi network environment for the target device.
This WiFi network works in 2.4G band with 20M bandwidth.
A Lenovo laptop is used in part of the data collection to provide sufficient data packets for audio digit classification by sending ICMP packets to the target device.
Devices used in our experiments are shown in the Figure \ref{fig:dev}.

For different digits, a volunteer record an approximately one-minute audio for each of the digits 0 through 9, respectively.
The device used to record the audio is the smartphone's own recorder app used in our experiments.
While the audio is being recorded, we ask the volunteer to say the number every two seconds.
Thus, the audio for one number contains the voice of the number 30 times.
We synthesize the recorded audios of the ten digits into \textit{a combined audio} in order to facilitate the collection of CSI data for all ten digits at once.
When the speaker is not playing sound, we mark it as \textit{silence} data.
Thus, we collect a total of 11 categories of data and used one-hot as their label.

\begin{table*}[!t]
	\centering
	\footnotesize
	\caption{Top-N accuracy for Each Digit and Silence}
	\begin{tabular}{m{1.3cm}<{\centering}m{0.6cm}<{\centering}m{0.6cm}<{\centering}m{0.6cm}<{\centering}m{0.6cm}<{\centering}m{0.6cm}<{\centering}m{0.6cm}<{\centering}m{0.6cm}<{\centering}m{0.6cm}<{\centering}m{0.6cm}<{\centering}m{0.6cm}<{\centering}m{0.7cm}<{\centering}m{0.7cm}<{\centering}}
		\toprule
		$P_N$   & 1 & 2 & 3 & 4  & 5 & 6 & 7 & 8 & 9  & 0  & Silence   \\ 
		\midrule
		$P_1$ & 16.67\% & 0 & 8.70\% & 10.71\% & 6.90\% & 28.57\% & 3.45\% & 0 & 100\% & 0 & 100\% \\
		$P_2$  & 33.33\% & 0 & 30.43\% & 32.14\% & 27.59\% & 57.14\% & 6.90\% & 4.76\% & 100\% & 5.26\% & 100\% \\
		$P_3$ & 58.33\% & 0 & 34.78\% & 53.57\% & 41.38\% & 71.43\% & 13.79\% & 14.29\% & 100\% & 15.79\% & 100\% \\
		$P_4$  & 66.67\% & 17.39\% & 52.17\% & 64.29\% & 48.28\% & 85.71\% & 24.14\% & 28.57\% & 100\% & 21.05\% & 100\% \\
		$P_5$ & 75.00\% & 21.74\% & 65.22\% & 75.00\% & 62.07\% & 92.86\% & 34.48\% & 28.57\% & 100\% & 36.84\% & 100\% \\
		\bottomrule
	\end{tabular}
	\label{tab:overall}
\end{table*}
We use the \textit{tcpdump} command to collect CSI data in the console app on the nexus 6P smartphone.
While collecting the data, we manually synchronize the target device to play the audio and the monitoring device to collect the CSI data.
Specifically, we manually press the audio playback key and the enter key of the command at the same time.
Additionally, due to CSI's sensitivity to the physical channel,
it captures the channel change of pressing the ENTER key.
Therefore, in the actual sample processing, we discard the first two seconds of the corresponding audio.
As for the monitor device, it cannot receive CSI packets stably due to its limitations.
The rated rate of ICMP packets sent by the laptop is 100Hz,
but the actual rate of packets collected by the listening device is between 60Hz and 150Hz.
The reason for lower than 100Hz is the existence of a certain packet loss rate, while the reason for higher than 100Hz is the data exchange between the target device and the router.
Therefore, we stabilize the number of CSI measurements in the sample at 200 according to the Algorithm \ref{alg1},
where $N_{norm}$ is set as 200.
When collecting data, the default target device, distance and volume are the honor v30 pro, 0.5 meters, and 60\% of the maximum volume, respectively.
We collect a total of about 4,862,200 CSI measurements.

We use Matlab R2020b to extract CSI from source files collect by monitor device.
The classification model is trained in Python 3.8.10 with the CUDA 11.4 library,
running on a machine with an Intel(R) Xeon(R) CPU E5-2680, 128G memory, and a Tesla P4 GPU.
For the TS-Net training, we use Adam optimizer with a learning rate of 0.001and the weight\_decay 0.03.
The parameter of the Dropout function is 0.5.

\subsection{Evaluation Metrics}
In this paper, we employ the top-N accuracy as the primary evaluation metric to evaluate the performance of our classification model.
In our experiment, this metric evaluates the accuracy of samples where the true label is present among the first N predictions generated by the model.
Mathematically, the top-N accuracy can be expressed as:
\begin{equation}
  P_N = \frac{1}{M} \sum_{i=1}^{M}\mathbb{I} (y_i \in \hat{Y}_i ^{N} ),
\end{equation}
where $M$ is the total number of samples in the test dataset, 
$y_i$ represents the true label for the $i_{th}$ sample,
and $\hat{Y}_i ^{N}$ denotes the set of the first $N$ predictions for the $i_{th}$ sample.
The indicator function $\mathbb{I}$ returns 1 if the true label is included in the first N predictions and 0 otherwise.

By focusing on top-N accuracy, we can better capture the model's ability to recover the digit contents in practical applications,
where users often prefer a list of likely candidates instead of a single identification.
This metric provides insights into the model's effectiveness in ranking potential labels and is particularly beneficial for tasks involving password predictions, search queries, and other scenarios requiring ranked predictions.
In our experiments, we mainly use top-5 accuracy $P_5$ to measure the performance of our model.

\subsection{Parameters Study}
To study the impact of the different parameters in the classification model,
We experimentally evaluate the accuracy of the model under different parameters.
First, we evaluate the impact of parameters $\alpha$ and $\beta$ in the Eq. \ref{eq:f} on the model accuracy.
Under the constraint of $\alpha + \beta = 1$, 
we set the value of $\alpha$ to 11 values with 0 as the initial value and 0.1 as the step size, respectively.
We train a model for each set of values of $\alpha$ and $\beta$, respectively.
Specifically, we collect CSI data while the target device plays the combined audio.
Under the default setting, we collect ten sets of CSI data and select a different set of CSI data as the test dataset for each training.

\begin{figure}[!t]
  \centering
  \subfloat[Parameter $\alpha$]{\includegraphics[width=0.45\linewidth]{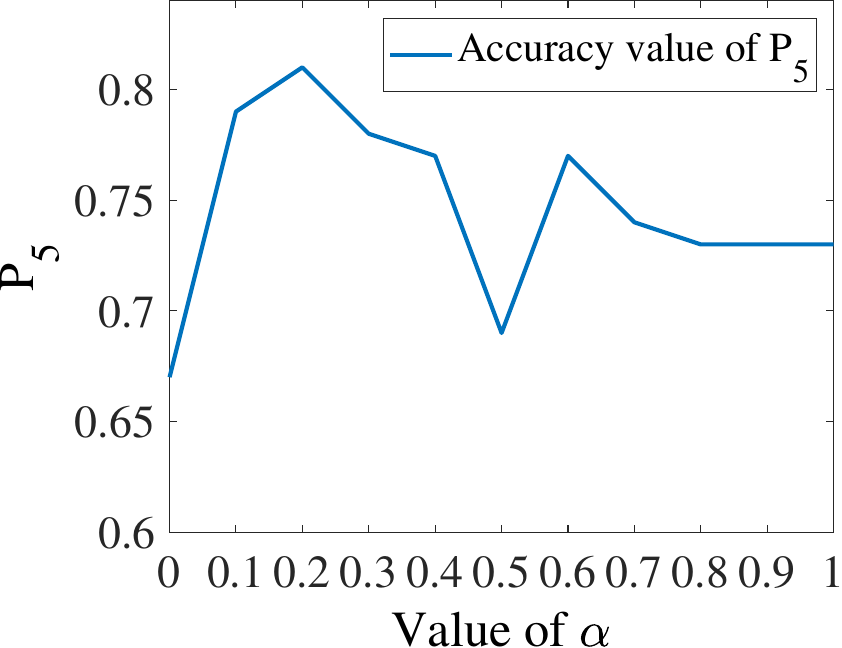}}
  \hspace{0.1em}
  \subfloat[Parameter epoch]{\includegraphics[width=0.45\linewidth]{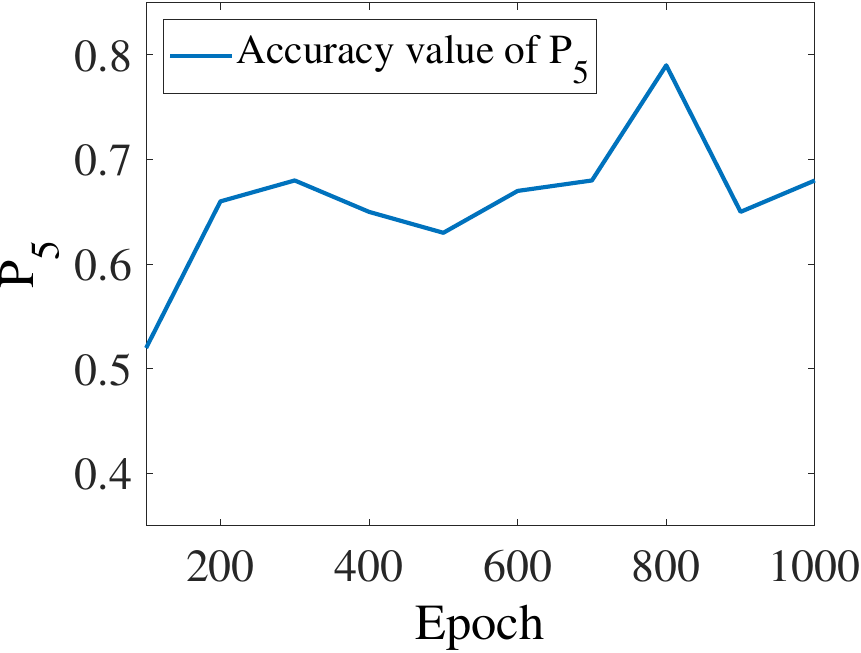}}
  \caption{Top-5 accuracy $P_5$ of the test dataset under different $\alpha$ values and epochs.}
  \label{fig:AlphValue}
\end{figure}
In this experiment, we employ the top-5 accuracy $P_5$ as the metric.
As shown in the Figure \ref{fig:AlphValue}(a),
as $\alpha$ increases, $P_5$ first rises and then falls, finally converging to a stable value 0.73.
When $\alpha$ equals 0.2, $P_5$ reaches a relatively high value 0.81.
And combining evaluations under other settings,
we determine the values of $\alpha$ and $\beta$ as 0.2 and 0.8 to obtain the good performance.

Next, we evaluate the impact of different training epochs on the accuracy.
In this study, we also use the same dataset in the last parameter study.
We set the training epoch to 11 values with 100 as the initial value and 100 as the step size.
The top-5 accuracy is shown in Figure \ref{fig:AlphValue}(b).
When the epoch equals 100, the top-5 accuracy is only 0.52.
As the epoch increases, the maximum value of the accuracies is 0.79 when the epoch is 800.
When the epoch is greater than 800, the accuracy decreases slightly,
probably because the model is overfitting.
Therefore, we select 800 as the value of the training epoch.

\subsection{Overall Performance}
\label{sec:4.4}
We study the overall performance of our system under the default setting and collect ten sets of CSI data.
Wei et al. \cite{WiVib} proposed a CSI-based loudspeaker sound recognition scheme like ours.
However, since our monitor device is only a smartphone configured with one antenna, its performance is much lower than the customized receiver configured with a 4$\times$4 antenna customized in \cite{WiVib}.
Our data does not provide enough and accurate phase information to realize their scheme.
However, our scheme is more realistic than \cite{WiVib}, revealing the possibility of a common smartphone eavesdropping on the sound emitted from other phones' loudspeakers.

As shown in Table \ref{tab:overall}, the top1 accuracy $P_1$ is very low, with individual digits (i.e., 2, 8, 0) having an accuracy of 0.
The accuracy of the two classes, the digit 9 and \textit{Silence}, has been maintained at 100\%.
As \textit{N} increases, the accuracy is increasing.
The average accuracy value for $P_1$ to $P_5$ are 24.99\%, 36.14\%, 45.76\%, 55.30\% and 62.89\%, respectively.
This indicates that our scheme can recognize the correct number in 5 guesses with high probability.
Therefore, we use top-5 accuracy $P_5$ as the primary evaluation metric in subsequent evaluations.

  \begin{figure}[!t]
	\centering
	\includegraphics[width=0.65\linewidth]{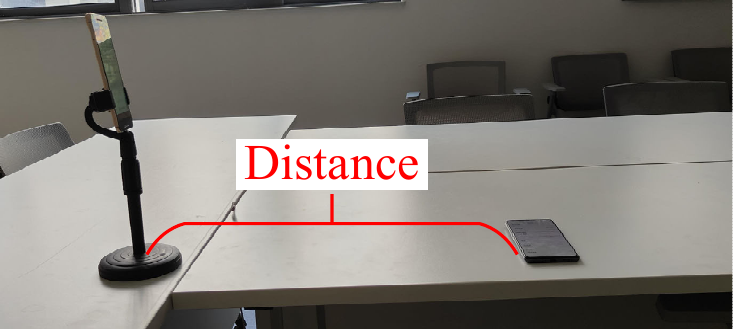}
    \caption{Schematic of distance between the monitor device and the target device.}
	\label{fig:1m}
\end{figure}

\subsection{Impact of Distance between Target Device and Monitor Device}
To evaluate the impact of the distance between monitor device and target device on the classification accuracy,
we collect CSI data by placing the monitor device at 5cm, 0.5m, 1m, 1.5m, 2m, 2.5m, 3m, 3.5m, 4m away from the target device (Honor V30 pro), as shown in Figure \ref{fig:1m}.
At each distance, we collect ten sets of CSI data while the target device plays the combined audio.

\begin{figure}[!t]
	\centering
	\includegraphics[width=0.57\linewidth]{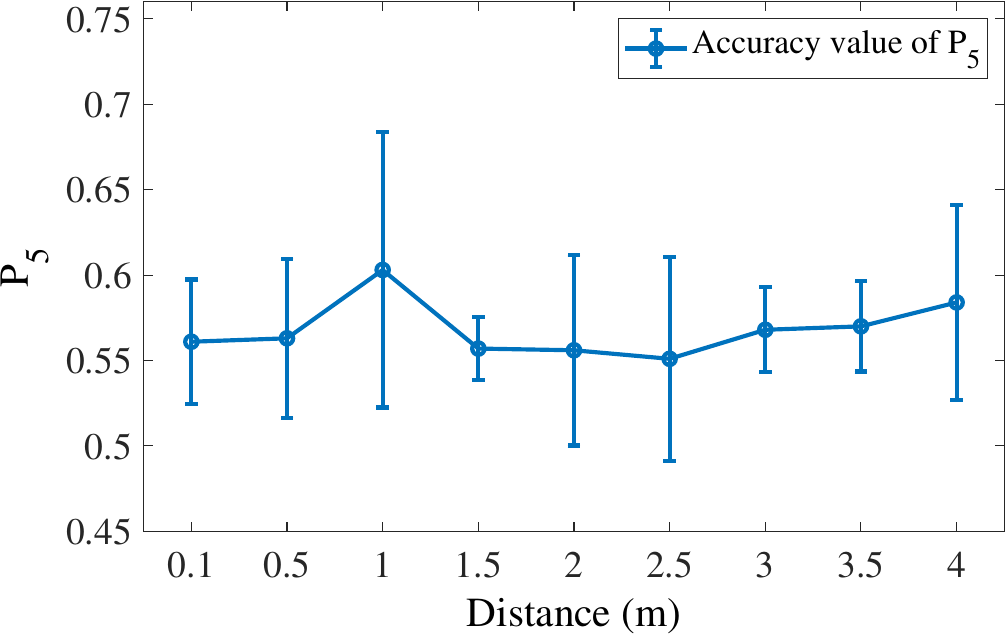}
	\caption{Top-5 accuracy $P_5$ of the test dataset under different distances.}
	\label{fig:distance}
  \end{figure}

We train a model with nine sets of CSI data at each distance.
The remaining set of data is used to test this model.
Theoretically, accuracy should decrease with distance.
However, as shown in Figure \ref{fig:distance}, as the distance increases,
the accuracy $P_5$ is increasing and then decreasing.
At the distance of 1 meter, the average accuracy is 60.3\%.
And after this distance, the accuracy of the model is slowly decreasing.
For example, at a distance of 4 meters, the accuracy still can reach 58.4\%.
First, regarding the phenomenon that the closest distance is not the most accurate,
our conjecture is that there is interference from the presence of the human body.
At the time of data collection, the human body is sitting near the target device.
A stationary human can also have an effect on CSI \cite{wifileaks,DeMan}.
Therefore, when the distance is close, the human body becomes the most significant influence on the CSI data from the monitor device.
And when the distance is gradually increased, the influence factors of the surrounding environment gradually increase.
Therefore, the influences at the source are gradually altered and cannot be recognized accurately.

\begin{table*}[!t]
	\centering
	\footnotesize
	\caption{$P_5$ of Different Distances under Different Distance Models}
	\begin{tabular}{m{1.5cm}<{\centering}m{0.8cm}<{\centering}m{0.8cm}<{\centering}m{0.8cm}<{\centering}m{0.8cm}<{\centering}m{0.8cm}<{\centering}m{0.8cm}<{\centering}m{0.8cm}<{\centering}m{0.8cm}<{\centering}m{0.8cm}<{\centering}}
		\toprule
		Model   & 5cm & 0.5m & 1m & 1.5m  & 2m & 2.5m & 3m & 3.5m & 4m     \\ 
		\midrule
		$Model_{5cm}$ & 53.05\% & 50.63\% & 61.26\% & 64.84\% & 56.67\% & 59.09\% & 58.99\% & 63.54\% & 52.58\%  \\
		$Model_{0.5m}$ &49.77\% & 55.94\% & 67.57\% & 62.64\% & 51.25\% & 55.05\% & 60.25\% & 57.46\% & 62.44\%  \\
		$Model_{1m}$ &52.58\% & 60\% & 72.97\% & 67.03\% & 55.83\% & 56.57\% & 58.04\% & 56.91\% & 62.44\%  \\
		$Model_{1.5m}$ &56.81\% & 55.94\% & 45.05\% & 54.95\% & 58.33\% & 55.05\% & 57.73\% & 64.64\% & 51.64\%  \\
		$Model_{2m}$ &51.64\% & 54.06\% & 61.26\% & 72.53\% & 53.33\% & 57.07\% & 56.15\% & 53.59\% & 59.15\%  \\
		$Model_{2.5m}$ & 53.99\% & 53.13\% & 74.78\% & 67.03\% & 61.25\% & 52.02\% & 55.52\% & 59.12\% & 61.03\% \\
		$Model_{3m}$ & 64.79\% & 54.69\% & 72.97\% & 69.23\% & 56.67\% & 51.52\% & 55.21\% & 60.22\% & 52.11\%  \\
		$Model_{3.5m}$ & 61.97\% & 55.94\% & 48.65\% & 72.53\% & 63.33\% & 56.06\% & 56.15\% & 59.67\% & 47.42\%  \\
		$Model_{4m}$ & 47.42\% & 54.38\% & 45.95\% & 57.14\% & 59.58\% & 53.54\% & 58.36\% & 60.22\% & 59.62\%  \\
		$Model_{all}$ & 65.26\% & 54.69\% & 70.27\% & 100\% & 67.92\% & 45.96\% & 55.21\% & 56.35\% & 54.46\%  \\
		\bottomrule
	\end{tabular}
	\label{tab:distance}
\end{table*}
To evaluate the generalization of different distance models,
we evaluate nine models trained with the CSI data collected at each distance.
As shown in Table \ref{tab:distance}, the accuracy at each distance is relatively low.
And the average accuracies are 57.85\%, 58.04\%, 60.26\%, 55.57\%, 57.64\%, 59.76\%, 59.71\%, 57.97\%and 55.13\%.
These results suggest that the models trained on individual distances lack generalization capability.
We conjecture that there may be ambient noise in the data that we have not filtered out.
The low signal-to-noise data makes the model does not learn useful knowledge to converge the loss to a stable value.
Moreover, considering a realistic scenario, it is cumbersome and unreasonable to train a separate model for the data at each distance.
Therefore, we use nine sets of data at all distances to train a model $Model_{all}$ at all distances,
leaving the remaining data as the test data set.
As shown in Table \ref{tab:distance}, the average accuracy is 63.35\%, which indicates that the features of the same digit sound vary somewhat at different distances.
Therefore, the effect of collecting data from different distances to improve the generalization of the model is limited.

\subsection{Impact of Target Devices' Diversity}
To verify the prevalence of this leakage, we evaluate our scheme using three smartphones as target devices.
Under the default conditions, we collect ten sets of CSI data for each target device.
Nine sets of CSI data are used to train a classification model for each device,
and last one set of CSI data is used to test the model.

\begin{figure}[!t]
	\centering
	\subfloat[Device]{\includegraphics[width=0.45\linewidth]{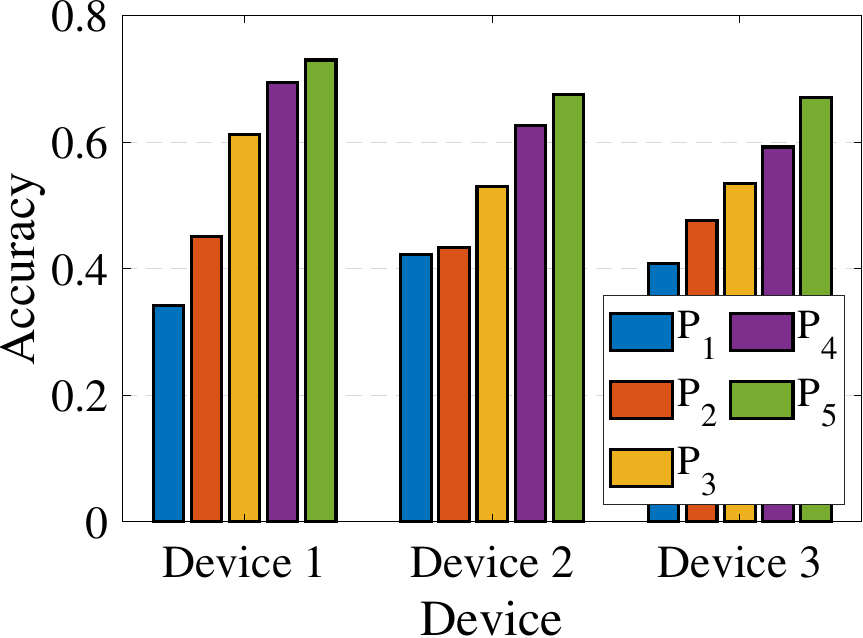}}
	\hspace{0.1em}
	\subfloat[Volume]{\includegraphics[width=0.45\linewidth]{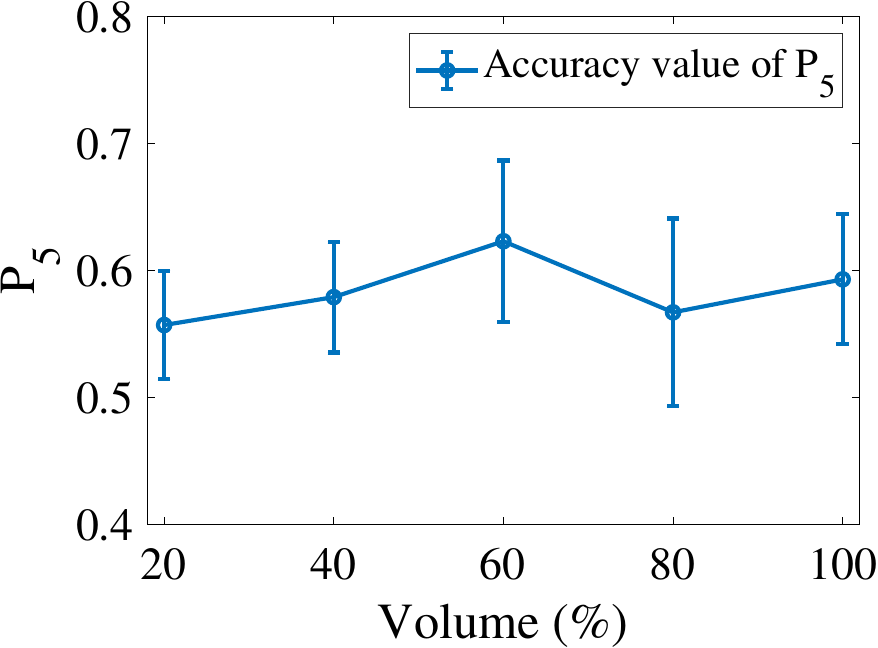}}
	\caption{Accuracy of the test dataset under different devices and volumes.}
	\label{fig:device}
  \end{figure}

The evaluation results are shown in Figure \ref{fig:device}(a).
The accuracy of the three devices is essentially the same.
Their $P_1$ and $P_5$ are 34.23\%, 42.17\%, 40.78\% and 72.97\%, 67.47\% and 66.99\%, respectively.
Even with $P_5$, the maximum accuracy is not 75\%.
This evaluation result indicates that the leakage of data from different devices is relatively small,
and it is not a significant threat to the model's performance.
In order to test the generalization of the model, we test the model with different devices using test data from different devices respectively.
The test results show that the value of $P_1$ is less than 20\%.
The training set contains data from two devices and the test set is from a third device.
The test results in this case are still less than 20\%.
These evaluation results suggest that collecting data from different devices can not improve the model's generalization ability. 
This indicates that the CSI mode is different for different devices.

  \subsection{Impact of Target Device's Volumes}
  To evaluate the impact of the device's volume on the model's performance,
  we set the target device's volume to 20\%, 40\%, 60\%, 80\%, and 100\%.
  At each volume level, the target device played the combined audio ten times.
  Similarly, we collect ten sets of CSI data for each volume and train classification models using the data from each volume level with leave-one-out method.
  We present the evaluation results of the models as shown in Figure \ref{fig:device}(b).
  The evaluation results show that the accuracy of the models trained with lower volumes is not necessarily lower than that of those trained with higher volumes.
  For instance, models corresponding to 40\% volume have an average accuracy of 57.9\%,
  while models trained with 80\% volume have an average accuracy of 56.7\%.
This result suggests that there is not a strong correlation between model accuracy and volume level.
The effects of vibration and the circuitry controlling volume on WiFi may not be captured by CSI, or their effects may not be learned during model training.

\subsection{Impact of Subject's Motion}
To evaluate the impact of the subject's movement on digit classification, we keep the target device stationary and collect CSI data under the moving condition of the subject.
Similarly, during each data collection, the target device plays the combined audio.
For the data collected when the subject is stationary, we test using the model trained in Section \ref{sec:4.4}.
For the data collected when the subject is moving, we train and test a model using the same method as before.
The average $P_5$ under the stationary condition is 72.97\% as discussed in Section \ref{sec:4.4}.
The model trained with moving data is almost unable to distinguish the played digits.
The average $P_5$ is only 23.51\%.
This result is predictable, due to that the impact of a moving subject on CSI is obviously greater than the electromagnetic leakage caused by the loudspeaker.
Many current works \cite{wifileaks,10379004,DeMan,zhu_et_2020} on movement detection and activity recognition also support this point.
Therefore, this evaluation result highlights the limitations of our attack.
It only achieves relatively high accuracy when the surrounding environment is relatively static.

  \begin{figure}[!t]
	\centering
	\includegraphics[width=0.8\linewidth]{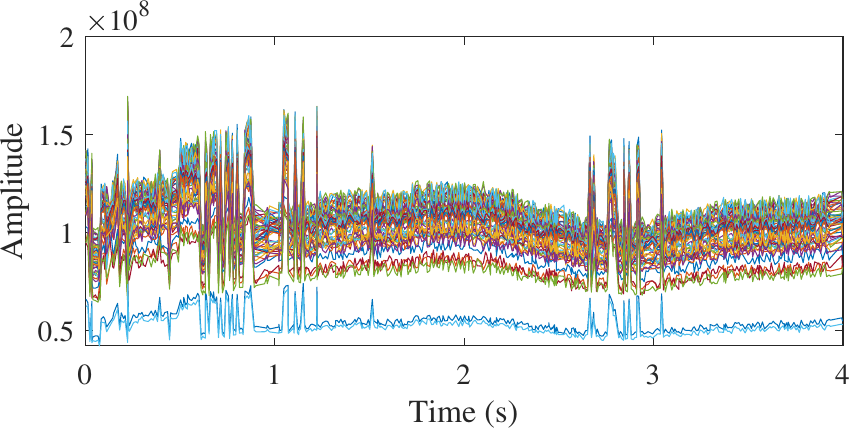}
    \caption{Amplitude of CSI subcarriers during playing the digit 7 twice.}
	\label{fig:obst}
\end{figure}

\begin{table*}[!t]
	\centering
	\footnotesize
	\caption{$P_1$ of Different Audio Digits under Different Models Trained with Different One-Minute Datasets}
	\begin{tabular}{m{1.5cm}<{\centering}m{0.7cm}<{\centering}m{0.7cm}<{\centering}m{0.7cm}<{\centering}m{0.7cm}<{\centering}m{0.7cm}<{\centering}m{0.7cm}<{\centering}m{0.7cm}<{\centering}m{0.7cm}<{\centering}m{0.7cm}<{\centering}m{0.7cm}<{\centering}m{0.7cm}<{\centering}}
		\toprule
		Model   & 1 & 2 & 3 & 4  & 5 & 6 & 7 & 8 & 9  & 0  & Silence    \\ 
		\midrule
		Model 1 & 92.59\% & 0 & 100\% & 50\% & 96.67\% & 58.62\% & 100\% & 100\% & 100\% & 100\% & 96.67\% \\
		Model 2 &74.07\% & 83.33\% & 62.96\% & 83.33\% & 100\% & 13.79\% & 100\% & 100\% & 86.67\% & 100\% & 93.33\% \\
		Model 3 & 0 & 93.33\% & 0 & 46.67\% & 100\% & 100\% & 100\% & 100\% & 100\% & 100\% & 93.33\% \\
		Model 4 & 39.29\% & 6.67\% & 14.82\% & 100\% & 100\% & 100\% & 100\% & 100\% & 100\% & 100\% & 93.33\% \\
		Model 5 & 35.71\% & 23.33\% & 100\% & 20\% & 100\% & 100\% & 100\% & 100\% & 100\% & 100\% & 93.33\% \\
		Model 6 & 53.57\% & 6.67\% & 92.59\% & 33.33\% & 86.67\% & 0 & 100\% & 100\% & 53.33\% & 100\% & 93.33\% \\
		Model 7 & 39.29\% & 0 & 96.30\% & 80\% & 100\% & 0 & 96.67\% & 100\% & 100\% & 100\% & 100\% \\
		Model 8 & 100\% & 3.33\% & 55.56\% & 16.67\% & 100\% & 3.45\% & 96.67\% & 100\% & 10\% & 86.21\% & 100\% \\
		Model 9 & 92.59\% & 0 & 44.44\% & 0 & 100\% & 100\% & 10\% & 100\% & 63.33\% & 100\% & 100\% \\
		Model 10 & 85.19\% & 0 & 44.44\% & 30\% & 100\% & 96.55\% & 13.33\% & 100\% & 100\% & 100\% & 100\% \\
		\bottomrule
	\end{tabular}
	\label{tab:num}
\end{table*}

\subsection{Classification with Obstacles}
To evaluate the system's performance under obstructions, we first place the monitor device in a cardboard box to block the line of sight.
Next, we place the monitor device outside the room, where the distance between the target device and the monitor device is approximately 1 meter, with a 5 cm thick wooden door between them.
Finally, we place the monitor device in the adjacent room, where these two devices are separated by a 30 cm thick concrete wall.
In all three scenarios, we played the combined audio and simultaneously collected CSI data.

According to the data segmentation method mentioned in Section \ref{sec:4.4}, we train a classification model for each of these three scenarios.
The first scenario has a $P_1$ accuracy of 23.74\%, while the other two scenarios are below 10\%.
This result shows that our scheme does not have the ability to go through walls.
Even in scenarios with simple obstacles, our scheme does not perform well.
This may be because our monitor device are too weak to acquire high-resolution signals.

\subsection{Low Accuracy Study}
\label{sec:4.9}
In the overall performance, we can find that the accuracy of top-1 is relatively low, only 34.23\%.
By analyzing the raw data, we find that not all CSI data can show a significant response to electromagnetic interference like Figure \ref{fig:csiempsev}.
As shown in the Figure \ref{fig:obst}, the difference ratio of CSI changes for the same number is large, and the second change is less and weaker.
The variability between similar data results in the model not being able to learn effective features.
As a result, the final trained model makes the classification accuracy relatively low.
As for the reason why CSI data does not produce correlated changes in loudspeaker playback, we speculate that it is due to the instability of electromagnetic radiation and the unstable sampling rate of the monitor device.

On the one hand, the loudspeaker's internal circuitry may generate spiky noise during digital-to-analog conversion and other signal processing.
Such noise spikes are typically high-frequency and have the potential to affect WiFi signals through electromagnetic coupling or conduction coupling.
However, this noise caused by playing audio does not always occur.
Therefore, not all of the CSI samples we collected will exhibit a strong correlation with audio playback.
On the other hand, our listening device is a smartphone configured with one antenna.
It does not acquire spatial and phase information with sufficient resolution.
Moreover, its sampling rate is not stable before an upper limit exists.
These limitations restrict the CSI characterization to have the interference generated by the loudspeaker.

By examining the timestamps of the CSI sequences, we find that the sampling rate of the received packets is variable.
This results in the samples obtained according to Algorithm 2, which are less in number than they should be, and many of them are obtained by linear interpolation in terms of quality.
This may be caused by the instability of the data collected by the monitoring device due to the long time of data collection.
Therefore, we only collected CSI data for one minute at a time.
We play the audio for each digit to collect the corresponding CSI data.
With the default settings, 10 one-minute pieces of data are collected for each digit.
We still use the leave-one-out method to train the model.
For each number of ten sets of data, we randomly select one set as test data and the other data as training data to train a model.
Following this method, we train a total of 10 different models.
The top-1 accuracy of the test data is shown in Table \ref{tab:num}.
The models are less stable, e.g., the accuracy of an individual model for a certain number is 0.
However, the overall average accuracy mostly can reach more than 75\%.
This result indicates that data collected for a single digit is better than data for all digits in terms of model accuracy and generalization.

\section{Discussions}
\label{sec:5}
In this section, we discuss countermeasures and limitations of our work,
and some additional findings.

\subsection{Potential Countermeasures}
While recovering digits from CSI is hardly a practical attack based on current system performance, we would still like to offer some potential countermeasures from both a device and CSI perspective.

From the perspective of the device itself, device vendors can optimize the packaging process to reduce electromagnetic leakage.
In particular, other sensor components such as loudspeakers, which were not considered before,
should be given additional electromagnetic protection for the corresponding components.
For example, the signal-to-noise ratio of electromagnetic signals can be reduced by means of a smartphone casing or by adding local electromagnetic shielding.
Another method is to plan the location of various components more rationally.
Previously, electronic devices such as smartphones do not seem to give sufficient consideration to possible electromagnetic interference from loudspeakers when the motherboard is designed.
Therefore, the electromagnetic interference between the WiFi antenna and the loudspeaker can be reduced by increasing the location of the two.

On the other hand, we can also defend against this potential attack from the perspective of CSI.
For example, Zhu et al. \cite{zhu_et_2020} employed a method that utilizes the router to send fake data packets with random power.
This method simulates WiFi data packets sent by terminal devices using the MAC address of the connected terminal device.
Since the monitor device filters packets based on the MAC address, it might collect WiFi packets with the same MAC address that are not sent by the same device.
Packets sent by different devices experience different channel changes, causing variations in the CSI distribution that the listener receives, making it difficult to classify them correctly.

\subsection{Limitations}
Although our scheme can recover audio digital content under certain conditions, we still have some limitations.

\textit{Root access.} The monitor device used in our work is a Nexus 6P smartphone.
Collecting CSI data requires root access on the phone.
Root access is quite sensitive and risky within the phone system.
Thus, it may pose additional privacy risks.
However, CSI extraction tools can also be deployed on embedded platforms such as Raspberry Pi.
Therefore, we are also exploring the construction of standalone devices to deploy our system.

\textit{Susceptibility to interference.} Due to the high sensitivity of CSI to channel changes,
the movement of objects in the vicinity of the target device can significantly impact the performance of our scheme.
Thus, CSI2Dig can currently achieve relatively high accuracy only in static environments and subjects.
This might be related to the small amount of data or the lack of high signal-to-noise ratio signals in data preprocessing.
We will further explore this issue in future work.

\textit{Continuous digit recognition.}
As can be seen in Section \ref{sec:4.4}, our scheme has a low $P_1$ value for collected digits.
Therefore, it can be conjectured that when we recognize a string of consecutive digits,
the probability of recognizing them correctly is very low according to the probabilistic calculus.
Although this can be mitigated by outputting multiple speculative results (e.g., $P_5$),
this is of very limited use for practical applications.
In Section \ref{sec:4.9}, it is possible to improve the accuracy by using CSI data collected over a short period of time to train the model.
However, when testing the model with data collected over a long period of time,
the accuracy $P_1$ is still less than 35\%.
Therefore, continuous digit recognition is a task that we are currently unable to accomplish.

\subsection{Arbitrary Acoustic Eavesdropping}
The essence of our scheme is to establish a corresponding CSI fingerprint for each digit played by the loudspeaker.
As mentioned in Section \ref{sec:2}, the amplitude variations in CSI caused by the same digit do not exhibit good similarity over time.
Therefore, we cannot recover the waveform of the sound as in previous works based on millimeter waves.
Furthermore, the sound waves from the speaker undergo significant distortion after being modulated twice by electromagnetic and CSI signals.
It is challenging to recover meaningful sound signals directly from changes in CSI signals.
Complex variables and defects in the equipment itself make it difficult to establish a meaningful mathematical model to represent this process.
However, relying on the inexplicability and data-driven nature of deep learning,
we are exploring the establishment of a deep learning model to accomplish this task.

\subsection{Impact of WiFi Packets Frequency}
Theoretically, the higher the sampling rate the more effective it is in portraying the electromagnetic interference on CSI.
However, due to the instability and unreliability of the sampling device, in the actual process of data collection, we found that the monitor device could not completely listen to the packets sent by the target device.
We set the rate of ICMP packets in steps of 50 to range from 50 to 500.
However, after setting the rate above 300, the rate of the received packets became very unstable, and the rate fluctuated roughly between 100 and 300.
This phenomenon does not allow us to explore the effect of higher sampling rates on system performance.
However, we also note that some other CSI data collection tools can achieve very high sampling rates (e.g., 1000 Hz).
Therefore, we are also exploring the system effects of using other devices for CSI acquisition.

\section{Related Work}
\label{sec:6}
Existing eavesdropping schemes for loudspeakers can be broadly categorized into motion sensor-based, optical sensor-based, and RF signal-based schemes, each utilizing different technologies.

Motion sensor-based schemes are primarily applied to mobile devices, such as smartphones and earphones.
These methods leverage sensors like accelerometers \cite{Spearphone,ba2020learning,hu_accear_2022,EarSpy} and gyroscopes \cite{gyrophone,ispyu,gao_practical_2023} to detect vibrations caused by the loudspeaker,
which are then correlated with the sound fluctuations to recover the content played.
However, these schemes typically assume the presence of malware on the device to access sensor data.
With increasing sensitivity around app permissions, such methods may become limited or infeasible in the future.

Optical sensor-based schemes mainly utilize optical equipment \cite{2200fps,277188,lidarphone,lidar} to detect vibrations in resonant objects caused by the sound from the loudspeaker to recover the audio content.
For example, Davis et al. \cite{2200fps} used a high-speed camera (2200 fps) to capture the vibrations of nearby objects,
while Lamphone \cite{277188} employed an electro-optical sensor to observe vibrations on a lamp's surface to recover audio content.
However, these schemes require a clear line of sight between the optical sensor and the vibrating object, as well as specialized, often costly equipment, making them impractical for general use.

RF-based schemes focus on using high-frequency signals to characterize vibrations induced by the loudspeaker.
These schemes mainly utilize millimeter waves \cite{mmEve,mmphase,mmecho,9413508,120mmwave,9796940,mmphone,mmDropper}, RFID signals \cite{9355596,RFspy,10.1145/3494975}, electromagnetic signals \cite{chen_eavesdropping_2024,liao_eavesdropping_2024,TEMPEST,magear}, and WiFi signals \cite{WiVib}, among other RF signals \cite{UWHear}.
These schemes typically require specialized equipment or complex hardware configurations.
For instance, RFSpy \cite{RFspy} relies on a software-defined radio to capture customized signal sources, enabling it to detect the vibrations of an RFID tag caused by a headset. 
Similarly, Wei et al. \cite{WiVib} employed a 4×4 antenna array placed at a specific location to capture the effects of loudspeaker-induced vibrations on the signal.
In another approach, Hu et al. \cite{mmecho} used a high-frequency Frequency-Modulated Continuous Wave (FMCW) radar to measure the vibrations of reverberant objects, which can then be used to recover sound information.
Chen et al. \cite{chen_eavesdropping_2024} demonstrated the potential of recovering sound from a loudspeaker by exploiting electromagnetic leakage from the amplification module.
However, this scheme requires an electromagnetic sensor closed to the target device and is dependent on the use of a headset.

In contrast to these schemes, our scheme uses only commercially available WiFi devices to recover digit content played through a smartphone loudspeaker.
Importantly, our scheme does not require specialized equipment such as antenna arrays or signal detection hardware.
Furthermore, our work enables the possibility of remotely capturing sound information from loudspeakers using WiFi devices,
significantly expanding the potential for practical applications without the need for costly or complex setups.

\section{Conclusion}
\label{sec:7}
In this paper, we have proposed CSI2Dig, a scheme designed to recover digit content played by smartphone loudspeakers based on channel state information.
Our key finding is that the audio played through smartphone speakers can influence the WiFi signals emitted by the device’s antenna.
By employing a denoising neural network and a feature extraction network TS-Net, we amplify the electromagnetic interference and effectively capture both temporal and spatial features from the CSI data.
We implement and test our scheme on commercial devices, conducting comprehensive experiments across various scenarios.
The experimental results demonstrate that CSI2Dig can consistently recover digital content with a high level of top-5 accuracy in diverse conditions.
We can achieve an average accuracy of 58.4\% at four meters using only CSI data.
We also discuss potential defenses when our scenario translates into an attack.


	\end{document}